\documentclass[twocolumn,aps,prx,10pt]{revtex4-2}

\usepackage{amsmath}
\usepackage{amssymb}
\usepackage{graphicx}
\usepackage{hyperref}
\usepackage{dsfont}
\usepackage{newtxtext}
\usepackage[varvw]{newtxmath}
\usepackage{bbm}

\hypersetup{
    colorlinks,
    linkcolor={blue},
    citecolor={blue},
    urlcolor={blue}
}

\graphicspath{{./}{./figs/}}

\newcommand{\rme}{\mathrm{e}}
\newcommand{\rmi}{\mathrm{i}}
\newcommand{\rmd}{\mathrm{d}}
\newcommand{\pd}{^{\phantom\dagger}}

\begin{document}

\title{%
Nesting instability of gapless U(1) spin liquids with spinon Fermi pockets in two dimensions
}

\author{Wilhelm G.\ F.\ Kr\"{u}ger}
\author{Lukas Janssen}

\affiliation{Institut f\"ur Theoretische Physik and W\"urzburg-Dresden Cluster of Excellence ct.qmat, TU Dresden, 01062 Dresden, Germany}

%%%%%%%%%%%%%%%%%%%%%%%%%%%%%%%%%%%%%%%%%%%%%%%%%%%%%%%%%%%%%%%%%%%%%%%
\begin{abstract}
Quantum spin liquids are exotic states of matter that may be realized in frustrated quantum magnets and feature fractionalized excitations and emergent gauge fields.
Here, we consider a gapless U(1) spin liquid with spinon Fermi pockets in two spatial dimensions.
Such a state appears to be the most promising candidate to describe the exotic field-induced behavior observed in numerical simulations of the antiferromagnetic Kitaev honeycomb model.
A similar such state may also be responsible for the recently-reported quantum oscillations of the thermal conductivity in the field-induced quantum paramagnetic phase of $\alpha$-RuCl$_3$.
We consider the regime close to a Lifshitz transition, at which the spinon Fermi pockets shrink to small circles around high-symmetry points in the Brillouin zone.
By employing renormalization group and mean-field arguments, we demonstrate that interactions lead to a gap opening in the spinon spectrum at low temperatures, which can be understood as a nesting instability of the spinon Fermi surface. This leads to proliferation of monopole operators of the emergent U(1) gauge field and confinement of spinons.
While signatures of fractionalization may be observable at finite temperatures, the gapless U(1) spin liquid state with nested spinon Fermi pockets is ultimately unstable at low temperatures towards a conventional long-range-ordered ground state, such as a valence bond solid.
Implications for Kitaev materials in external magnetic fields are discussed.
\end{abstract}
%%%%%%%%%%%%%%%%%%%%%%%%%%%%%%%%%%%%%%%%%%%%%%%%%%%%%%%%%%%%%%%%%%%%%%%

\date{\today}

\maketitle

%%%%%%%%%%%%%%%%%%%%%%%%%%%%%%%%%%%%
\section{Introduction}
%%%%%%%%%%%%%%%%%%%%%%%%%%%%%%%%%%%%

Fractionalization is a fascinating many-body phenomenon that can occur in strongly-correlated electron systems.
An important example is given by frustrated quantum magnets with localized electrons~\cite{savary17, broholm20}.
Magnetic insulators with sizable charge gaps are usually described in terms of effective spin models. In conventional systems, the low-energy degrees of freedom are spin waves with magnons as the associated collective quasiparticle excitations.
In strongly-frustrated quantum magnets, however, the spin degrees of freedom can fractionalize into novel quasiparticles that interact via emergent gauge fields.
A paradigmatic theoretical example of this intriguing phenomenon is the Kitaev honeycomb model~\cite{kitaev06}. This model features a nearest-neighbor Ising-type interaction between spins-$1/2$ on the honeycomb lattice, with a direction-dependent spin quantization axis. It is this bond dependence of the exchange interaction that leads to spin frustration~\cite{trebst17}.

A number of materials in the strong-spin-orbit-coupled regime have been proposed to realize a sizable Kitaev exchange~\cite{winter17b}. Among these, the $d^5$~honeycomb magnets $A_2$IrO$_3$ with $A = \text{Li}, \text{Na}$~\cite{%
%initial:
chaloupka10, 
%experiment:
singh10, singh12, choi12, chun15, williams16, choi19, 
%theory:
foyevtsova13, katukuri14, sizyuk14, sizyuk16, winter16},
their hydrogen intercalated modification H$_3$LiIr$_2$O$_6$~\cite{%
kitagawa18, yadav18, li18},
and, in particular, $\alpha$-RuCl$_3$~\cite{%
%experiments:
plumb14, sears15, sears17, sears20, johnson15, ran17, do17, banerjee16, banerjee17,
%theory:
yadav16, janssen17b, winter17a, janssen20a, andrade20} 
have received significant attention. 
Recent works have also suggested cobalt-based  $d^7$~honeycomb magnets, such as $A_2$Co$_2$TeO$_6$ and $A_3$Co$_2$SbO$_6$~\cite{liu18, sano18, songvilay20, vivanco20, chen21, hong21}, or BaCo$_2$(AsO$_4$)$_2$~\cite{zhong20, zhang21}, as promising candidates for the realization of the Kitaev model.
In the above examples, the geometry of edge-sharing octahedra surrounding the magnetic ions lead to a suppression of the nearest-neighbor Heisenberg interaction, leaving behind a dominant Kitaev exchange that stems from Hund's coupling mediated through the excited levels~\cite{jackeli09, trebst17}.
While all these materials (with the exception of H$_3$LiIr$_2$O$_6$~\cite{kitagawa18}) display magnetic order at the lowest temperatures, various experimental findings suggest that they are located in proximity to a genuine spin-liquid phase~\cite{singh12, chun15, nasu16, banerjee16, do17}.
Moreover, in $\alpha$-RuCl$_3$, the long-range order can be suppressed by a moderate in-plane magnetic field of around 7\,T, giving way to an exotic quantum paramagnetic state. The nature of this state has been a matter of intense debate~\cite{%
janssen19,
baek17, wolter17, gass20, hentrich18, hentrich19, hentrich20, banerjee18, balz19, balz21, bachus20, bachus21,
leahy17, zheng17, jansa18,
winter18, kaib19, chern21}.
Two of the most astonishing experimental findings in this state are a half-integer thermal Hall effect~\cite{kasahara18, yokoi20, bruin21}, suggesting the presence of gapless Majorana edge modes~\cite{vinkleraviv18, ye18}, and characteristic quantum oscillations in the longitudinal heat conductivity at low temperatures in a finite field range below 11\,T~\cite{czajka21}, indicative of a field-induced spin-liquid state with a spinon Fermi surface.
While the thermal Hall effect vanishes for field directions along Ru-Ru bonds~\cite{yokoi20}, quantum oscillations can be observed for all measured in-plane field directions~\cite{czajka21}. This peculiar feature of the experiments has recently been rationalized within a model of small isolated spinon Fermi pockets~\cite{sodemann21}.
In all of the above-mentioned materials, the Kitaev interaction is believed to be ferromagnetic~\cite{winter16, janssen17b, koitzsch20, sears20}.
More recently, materials with a dominant antiferromagnetic Kitaev interaction have also been suggested. This includes the $f$-electron based honeycomb magnets $A_2$PrO$_3$ with $A = \text{Li}, \text{Na}$~\cite{jang19}, as well as potential higher-spin realizations, such as $A_3$Ni$_2X$O$_6$ with $X = \text{Bi}, \text{Sb}$~\cite{stavropoulos19}.

In the absence of an external magnetic field, the Kitaev model is exactly solvable. In the isotropic limit, in which the strengths of the Kitaev couplings are equal on all bonds, the low-energy excitations are described by gapless Majorana fermions in the background of a gapped and static $\mathbbm Z_2$ gauge field. The system realizes a gapless $\mathbbm Z_2$ quantum spin liquid.
Applying a small external magnetic field to the system along a direction that does not happen to be perpendicular to one of the cubic spin quantization axes gaps out the Majorana spectrum and gives rise to non-Abelian anyon excitations above a topologically-ordered ground state~\cite{kitaev06}.
A finite magnetic field leads to fluctuations of the $\mathbbm Z_2$ gauge field, spoiling the exact solubility of the model.
The phase diagram of the model as a function of magnetic field has recently been vividly debated~\cite{janssen19}.
As can be observed already on the semiclassical level~\cite{janssen16b, consoli20}, the model with antiferromagnetic coupling features, in comparison with the ferromagnetic case, a significantly larger field range that could accommodate nontrivial field-induced phases:
In the ferromagnetic model, a very small field on the order of just a few percent of the Kitaev coupling is sufficient to drive a transition towards the high-field conventional paramagnet that is adiabatically connected to the fully-polarized state~\cite{jiang11}.
In the antiferromagnetic model, this conventional paramagnet is stabilized only at much larger field strengths~\cite{janssen16b,consoli20}.
For moderate fields, numerical studies have found indications for a novel field-induced intermediate phase, which is distinct from both the low-field topologically-ordered phase and the high-field conventional paramagnet~\cite{gohlke18, zhu18}.
Finite-size analyses of exact diagonalization results on clusters with up to 32 sites~\cite{hickey19, kaib19} and density matrix renormalization group calculations on cylinders with a circumference of up to five unit cells~\cite{zhu18, jiang18, jiang19, patel19} suggest that an exotic quantum paramagnet is realized in this intermediate-field regime.
A recent study that approaches the question from the limit of the Kitaev-Gamma ladder has come to a similar conclusion~\cite{soerensen21}.
Among the candidate states that are in proximity to the Kitaev spin liquid, the ones that appear most consistent with the numerical data for entanglement entropy~\cite{jiang18, jiang19} and static spin structure factor~\cite{patel19} are gapless U(1) spin liquids with a finite spinon Fermi surface that is divided into small pockets located around high-symmetry points in the Brillouin zone.
For a large field range, the pockets exhibit perfect~\cite{patel19} or approximate~\cite{jiang18} nesting, and degenerate to perfectly nested isolated Fermi points upon increasing the field strength towards the high-field transition, beyond which the conventional paramagnetic state is stabilized.
\begin{figure}[tb]
\includegraphics[width=\linewidth]{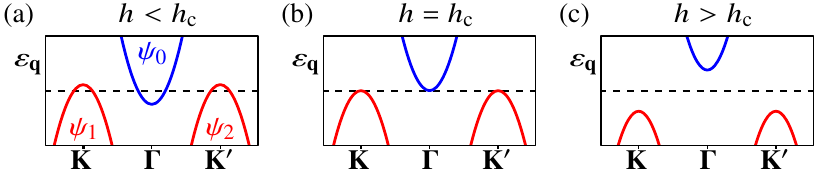}
\caption{(a) Schematic spinon band structure of U(1) spin liquid state with particle-like pocket around the center $\boldsymbol\Gamma$ of the hexagonal Brillouin zone and hole-like pockets around its corners $\mathbf K$ and $\mathbf K'$, proposed as effective description of the field-induced exotic quantum paramagnet occurring for intermediate field strengths $h < h_\mathrm{c}$ below the high-field transition in the antiferromagnetic Kitaev model~\cite{jiang18, jiang19}. The Fermi level is indicated as dashed line.
(b) Increasing the field strength shifts the particle-like (hole-like) band(s) up (down) in energy, such that for $h = h_\mathrm{c}$, the bands just touch the Fermi level at isolated points $\mathbf K$, $\mathbf K'$, and $\boldsymbol\Gamma$.
(c) In the high-field phase, the spinon spectrum is gapped, leading to proliferation of monopoles of the compact U(1) gauge field and confinement of spinons. This describes the conventional paramagnet that is adiabatically connected to the polarized state.}
\label{fig:bandstructure}
\end{figure}
In this scenario, the U(1) spin liquid can be understood as a parent state, out of which the non-Abelian Kitaev spin liquid at low fields emerges from pair condensation of spinons for sufficently strong attractive interactions between equally-charged spinons~\cite{hickey19}. The spinon condensate breaks the local U(1) symmetry down to a $\mathbbm Z_2$ subgroup and gaps out the gauge field via the Higgs mechanism~\cite{metlitski15}.
The conventional paramagnetic state at high fields arises from a Lifshitz transition at a critical field strength $h_\mathrm{c}$, at which the Fermi pockets shrink to isolated Fermi points located at high-symmetry wavevectors in the Brillouin zone. For larger field strengths $h>h_\mathrm{c}$, the spinon spectrum acquires a full gap, leading to proliferation of monopoles of the compact U(1) gauge field and confinement of spinons~\cite{polyakov75, polyakov77}, and resulting in a trivially gapped phase.
An example with a particle-like pocket at the center of the hexagonal Brillouin zone and two hole-like pockets near its corners is illustrated in Fig.~\ref{fig:bandstructure}.
We note that while the Kitaev interaction in $\alpha$-RuCl$_3$ is likely ferromagnetic~\cite{winter16, janssen17b, koitzsch20, sears20}, a dual version~\cite{chaloupka15} of the putative field-induced U(1) spin liquid can also appear in models relevant for this material, with a sizable ferromagnetic Kitaev coupling and additional Heisenberg and off-diagonal Gamma interactions~\cite{kaib19}.

\begin{figure}[tb]
\centering
\includegraphics[width=\linewidth]{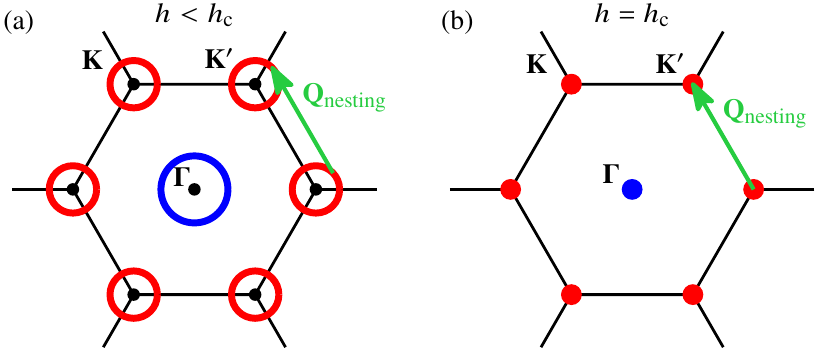}
\caption{(a) Nested spinon Fermi surface of U(1) spin liquid state for $h<h_\mathrm{c}$ with particle-like and hole-like pockets around $\boldsymbol\Gamma$ and $\mathbf K$, $\mathbf K'$ points in the hexagonal Brillouin zone, depicted as blue and red circles, respectively. The nesting vector $\mathbf Q_\text{nesting} = \mathbf{K}$ is indicated as green arrow.
(b) Upon approaching the Lifshitz transition at $h = h_\mathrm{c}$, the spinon Fermi pockets shrink to isolated spinon Fermi points at $\boldsymbol\Gamma$ and $\mathbf K$, $\mathbf K'$, depicted as blue and red dots, respectively.}
\label{fig:fermi-pockets}
\end{figure}

In this work, we investigate the stability of gapless U(1) spin liquids with nested spinon Fermi pockets in two spatial dimensions at low temperatures.
In order to gain theoretical control within a field-theoretical analysis, we focus on the regime close to a Lifshitz transition, in the vicinity of which the Fermi pockets are small, such as in Fig.~\ref{fig:bandstructure}(b).
We show that fluctuations of the gapless U(1) gauge field mediate a repulsive interaction between equally-charged spinons, leading to a suppression of spinon pairing. As a consequence, weak attractive interactions are irrelevant in the renormalization group (RG) sense and a pairing instability can only occur at sufficiently strong attractive interactions. This aspect is similar to the situation in a U(1) spin liquid with an extended and simply-connected spinon Fermi surface~\cite{metlitski15}.
However, in contrast to this latter situation, in the present case, the nesting of the Fermi pockets, see Fig.~\ref{fig:fermi-pockets}, renders certain repulsive spinon interactions RG relevant upon the inclusion of gauge-field fluctuations.
This leads to a full gap opening in the spinon spectrum and an instability of the U(1) spin liquid state at low temperatures.
We argue that the resulting new ground state is characterized by confinement of spinons and realizes a conventional long-range-ordered state, such as a valence bond solid (VBS)~\cite{read90, fouet01, ganesh13b, zhu13}.
The scenario that we propose for U(1) spin liquids with nested spinon Fermi pockets is similar to the situation of electronic quasiparticles in iron pnictides and chalcogenides~\cite{paglione10}. The normal state of this class of effectively two-dimensional materials is characterized by isolated Fermi pockets around high-symmetry points in the square Brillouin zone, and features approximate nesting. Interpocket interactions can drive nematic, magnetic, and exotic superconducting instabilities of the metallic state~\cite{fernandes14}. This can be understood already in the idealized limit of perfect nesting, in which the Fermi pockets shrink to isolated Fermi points at which the electron and hole bands just touch the Fermi level~\cite{chubukov12, fernandes17}.
The particular type of instability depends on microscopic parameters, such as doping level, sign and relative strength of interactions, shape of Fermi pockets, as well as temperature. 
In the present case of nested \emph{spinon} Fermi pockets, the main difference to the situation in the iron-based materials is the emergence of the $(2+1)$-dimensional U(1) gauge field, which, as we show, further enhances repulsive interpocket interactions between equally-charged Fermi excitations and suppresses pairing instabilities.

The remainder of this work is organized as follows: In Sec.~\ref{sec:models}, we introduce two minimal effective models describing U(1) spin liquid states with spinon Fermi pockets in the limit of perfect nesting. Section~\ref{sec:RG} contains a discussion of the RG flow of our models, revealing that gauge fluctuations effectively give rise to repulsive interpocket interactions and ultimately render the gapless spin liquid state unstable at low temperatures. In Sec~\ref{sec:MFT}, we argue that the low-temperature phase is characterized by a gapped spinon spectrum. Proliferation of monopole operators of the compact U(1) gauge field leads to confinement of spinons and a conventional long-range-ordered ground state, the properties of which are analyzed in Sec.~\ref{sec:discussion}. We present our conclusions, together with some comments on effects beyond our minimal modeling and a discussion of implications of our results for Kitaev materials in a magnetic field, in Sec.~\ref{sec:conclusions}. The appendix contains technical details of the RG calculation.

%%%%%%%%%%%%%%%%%%%%%%%%%%%%%%%%%%%%
\section{Effective models}
\label{sec:models}
%%%%%%%%%%%%%%%%%%%%%%%%%%%%%%%%%%%%

The spinons discussed in this work can be thought of as fermionic degrees of freedom that arise from fractionalization of spin-$1/2$ operators $\mathbf S_i$ at lattice sites $i$.
Formally, they can be understood from an Abrikosov decomposition~\cite{abrikosov65}
\begin{align} \label{eq:decomposition}
    \mathbf S_i = \frac{1}{2} \Psi^\dagger_i \boldsymbol\sigma \Psi\pd_i,
\end{align}
where $\Psi_i = (\Psi_{i\uparrow}, \Psi_{i\downarrow})^\top$ and $\Psi^\dagger_i = (\Psi^\dagger_{i\uparrow}, \Psi^\dagger_{i\downarrow})$ are two-component spinors representing the spinon degree of freedom and $\boldsymbol\sigma$ is the vector of Pauli matrices.
This construction obeys the usual SU(2) spin algebra provided that the spinon fields satisfy canonical anticommutation relations.
In general, however, the construction introduces fictitious states in which the particle number $\Psi_i^\dagger \Psi\pd_i$ at a given site $i$ is zero or two, thereby enlarging the local Hilbert space dimension from two to four. These unphysical states can be excluded by applying the half-filling constraint
\begin{align} \label{eq:constraint}
    \Psi^\dagger_i \Psi\pd_i \equiv 1
\end{align}
at each lattice site $i$, which projects the four-dimensional local Hilbert space to the two-dimensional singly-occupied subspace.
The decomposition in Eq.~\eqref{eq:decomposition} and the constraint in Eq.~\eqref{eq:constraint} feature an explicit invariance under local U(1) gauge transformations $\Psi_i \mapsto \rme^{\rmi \lambda_i} \Psi_i$. The full gauge redundancy is actually SU(2)~\cite{affleck88b, wen02}.

We assume a situation as in the Kitaev model in an external magnetic field, in which the degeneracy between up and down components of $\Psi$ is lifted by strong spin-orbit coupling and finite field.
Within a mean-field picture, this leads to four generically nondegenerate spinon bands on a lattice with two-site unit cell, such as the honeycomb lattice~\cite{zou18}.
Which ones of these spinon bands cross the Fermi level, and where in the Brillouin zone, depends on the symmetry of the system and the microscopic parameters of the original spin model.
In this work, we consider the situation in which there are small nondegenerate particle-like and hole-like spinon Fermi pockets around isolated points in the Brillouin zone.
This implies that two of the four bands are fully gapped and do not contribute to the low-energy physics.
We furthermore assume that Luttinger's theorem~\cite{luttinger60} holds for the spinon bands. The total area enclosed by the particle-like pockets must then be equal to the total area enclosed by the hole-like pockets.
In the following, we will explicitly study two minimal effective models describing such a situation.
As a warm-up, we consider a simple two-pocket model, which has been suggested as an effective low-energy description for a quadruple point separating the gapped non-Abelian and Abelian $\mathbb Z_2$ spin liquids, the gapless U(1) spin liquid, and the high-field paramagnetic phase in the anisotropic Kitaev honeycomb model in an external magnetic field along the $[111]$ direction~\cite{jiang18}.
We then investigate a three-pocket model with one particle-like spinon band and two hole-like spinon bands near the Fermi level. Such a model was proposed as an effective low-energy theory for the putative gapless U(1) spin liquid state in the isotropic Kitaev model with equal antiferromagnetic coupling strengths on all bonds in a $[111]$ field~\cite{jiang18, jiang19}.
A discussion of more elaborate models, including the four-pocket model that represents an alternative low-energy candidate theory for the same situation~\cite{patel19}, as well as of effects beyond the limit of perfect nesting, will be postponed to Sec.~\ref{sec:conclusions}.

%%%%%%%%%%%%%%%%%%%%%%%%%%%%%%%%%%%%
\subsection{Two-pocket model}
%%%%%%%%%%%%%%%%%%%%%%%%%%%%%%%%%%%%

The spinon Fermi surface in the two-pocket model consists of one particle-like pocket and one hole-like pocket in the Brillouin zone. 
In the realization proposed for the anisotropic Kitaev model~\cite{jiang18}, the particle-like pocket is located around the zone center $\boldsymbol\Gamma$, while the hole-like pocket resides around the midpoint $\mathbf M$ of one of the zone edges.
A minimal description of such a situation is given by the continuum action $S_2 = \int \rmd^2\mathbf x \rmd \tau \mathcal L_2$ with
\begin{equation}
\begin{split}
\mathcal{L}_{2}&=
\psi_0^\dagger \left[\partial_\tau-\rmi ea_\tau-\frac{m_1}{m_0}\tilde{\mu}+\frac{(-\rmi\boldsymbol\nabla-g\mathbf{a})^2}{2m_0}\right]\psi_0\pd\\
&\quad
+\psi_1^\dagger \left[\partial_\tau-\rmi ea_\tau+\tilde{\mu}-\frac{(-\rmi\boldsymbol\nabla-g\mathbf{a})^2}{2m_1}\right]\psi_1\pd\\
&\quad
+\frac{1}{4}f_{\mu\nu}f^{\mu\nu}
+u\psi_0^\dagger\psi_0\pd\psi_1^\dagger\psi_1\pd,
\end{split}
\label{eq:ansio_theory}
\end{equation}
where $\tau$ denotes imaginary time, and $\psi_0$ and $\psi_1$ are Grassmann fields describing spinon excitations in the particle-like and hole-like pockets, respectively. 
The parameters $m_0$ and $m_1$ denote the corresponding effective band masses, which we assume to be isotropic within each pocket for simplicity.
The spinons are minimally coupled to the components $a_\tau$ and $\mathbf a = (a_x, a_y)$ of the U(1) gauge field, with electric and magnetic couplings $e$ and $g$, respectively. In the above and the following equations, the summation convention over repeated space-time indices $\mu,\nu \in \{ 0,1,2\}$ is assumed.
Importantly, the model lacks Lorentz invariance, such that the electric scalar potential $a_\tau$ and the magnetic vector potential $\mathbf a$ are not related by symmetry and the electric and magnetic gauge couplings $e$ and $g$ are in general independent.
The antisymmetric real field strength tensor $f_{\mu\nu}$ is related to $a_\tau$ and $\mathbf a$ in the usual way,
\begin{equation}
\left(f_{\mu\nu}\right)=
\begin{pmatrix}0&\frac{\partial_\tau a_x-\partial_x a_\tau}{c}&\frac{\partial_\tau a_y - \partial_y a_\tau}{c}\\
\frac{\partial_x a_\tau-\partial_\tau a_x}{c}&0&\partial_x a_y - \partial_y a_x \\
\frac{\partial_y a_\tau-\partial_\tau a_y}{c}&\partial_y a_x - \partial_x a_y&0
\end{pmatrix},
\end{equation}
where $c>0$ is the speed of the ``artificial light''~\cite{wen03}.
As we demonstrate below, the coupling of gapless spinons to the gapless gauge field generates a local four-fermion interaction of the form $\psi_0^\dagger\psi_0\pd \psi_1^\dagger\psi_1\pd$, which has therefore been included in Eq.~\eqref{eq:ansio_theory} from the outset, with coupling parameter $u$.
It can be understood as a particle-hole interpocket interaction between the particle density $\psi_0^\dagger\psi_0\pd$ and the hole density $\psi_1^\dagger\psi_1\pd$, with $u>0$ ($u<0$) corresponding to attraction (repulsion) between oppositely-charged particles and holes.
The global chemical potential is fixed to comply with the half-filling constraint, Eq.~\eqref{eq:constraint}.
The parameter $\tilde{\mu}$ can be understood as a local chemical potential, allowing one to tune the sizes of the spinon Fermi pockets present for $\tilde{\mu}>0$ in the noninteracting limit. 
The prefactor $m_1/m_0$ in the $\tilde{\mu}$ term in the first line of Eq.~\eqref{eq:ansio_theory} ensures that particle-like and hole-like pockets enclose equal areas, as required by Luttinger's theorem for half filling.
Decreasing $\tilde{\mu}$ shifts the particle-like band up in energy and the hole-like band down, leading to a simultaneous shrinking of both pockets. For $\tilde{\mu} = 0$, the Fermi pockets degenerate to isolated Fermi points. Eventually, for $\tilde{\mu} < 0$, the spinon spectrum is fully gapped.
In Eq.~\eqref{eq:ansio_theory}, we have assumed that the U(1) gauge field is noncompact, a property that is emergent in the case of a finite spinon Fermi surface~\cite{lee08}, i.e., for $\tilde \mu > 0$.
However, when the spinon spectrum is gapped out, magnetic monopoles of the gauge field start to proliferate and lead to a gap in the gauge-field excitation spectrum and the confinement of spinons~\cite{polyakov75, polyakov77, hermele04}. This will be discussed in Sec.~\ref{sec:discussion}.
The physics of magnetic monopoles in the case of $\tilde \mu = 0$ requires a more detailed discussion. First of all, we note that previous works on monopoles in related (2+1)-dimensional gauge field theories~\cite{hermele04, lee08, pufu14, karthik19, xu19, janssen20b, dupuis19, chester21} suggest an intimate relationship between the monopole scaling dimension and the spinon density of states near the Fermi level: Typically, the larger the density of states, the larger the scaling dimension of the lowest monopole operator. This general rule of thumb is consistent with the finding that monopoles are irrelevant in the case of a finite Fermi surface~\cite{lee08} and relevant for a gapped spinon spectrum~\cite{polyakov75, polyakov77}. The marginal case occurs for Dirac spin liquids, in which case the single-particle density of states vanishes linearly at the Fermi level. For a large number of fermion flavors, all monopole operators are irrelevant~\cite{hermele04}, while for a small number of flavors below a certain ``critical'' flavor number, the lowest monopole operator becomes relevant~\cite{pufu14, karthik19}. For $\tilde\mu = 0$ in our models, the single-particle density of states remains finite at the Fermi level, and appearing thus above the ``critical'' density of states, suggesting that the irrelevance of monopoles for $\mu > 0$ remains true in the limit $\mu \to 0$, as long as the spinon spectrum remains gapless.

In the following sections, we consider the regime of small nonnegative $\tilde{\mu}$, in which the finite extents of the Fermi pockets can be neglected. This corresponds to the limit of perfect nesting, which is sufficient to understand the behavior of the system at energy scales small compared to the spinon bandwidth, but above the scale set by $\tilde{\mu}$~\cite{chubukov12, fernandes17}.
Importantly, this energy window allows a fully controlled RG approach, provided that the gauge field can effectively be considered as noncompact.
For small $\tilde{\mu}$, the effective band dispersion is $\varepsilon^{(0)}_{\mathbf q} = +\mathbf q^2/(2m_0)$ and $\varepsilon^{(1)}_{\mathbf q} = - \mathbf q^2/(2m_1)$, where $\mathbf q$ denotes the deviation from the respective Fermi point in the Brillouin zone.
At tree level, frequencies and momenta hence scale as $\omega \propto q^z$ with dynamical exponent $z=2$.
The power-counting inverse-length dimensions of the fields are consequently
\begin{align}
    [\psi_{0}] & = [\psi_{1}] = \frac{d}{2}, &
    [a_\tau] & = \frac{d+2}{2}, &
    [\mathbf a] & = \frac{d}{2}, 
\end{align}
and those of the couplings become
\begin{align} \label{eq:scaling-dims}
    [e] & = [g] = \frac{2-d}{2}, &
    [u] & = 2-d.
\end{align}
Importantly, all three couplings are simultaneously marginal in the physical dimension $d=2$, enabling a controlled perturbative analysis.
Since $z=2$ at tree level, the speed of artificial light $c$ has inverse length dimension $[c] = 1$, implying that $c$ is a relevant parameter and needs to be taken into account at all orders in the loop expansion.
One of the band masses, say $m_0$, can be absorbed by a rescaling of the fermion fields, leaving behind a single dimensionless mass imbalance ratio $m_1/m_0$.
As an aside, we note that an alternative RG scheme that fixes the dynamical exponent $z=1$, such that $c$ is a marginal parameter, is in principle possible. Such a scheme, however, would render the gauge couplings relevant at tree level, inhibiting a controlled perturbative analysis.
Within our scheme, the Lagrangian~\eqref{eq:ansio_theory} contains all marginal and relevant terms compatible with the field content and the symmetries of the model for small $\tilde \mu \geq 0$, and as such represents an appropriate starting point for the RG analysis.
For sizable $\tilde \mu >0$, when the spinon Fermi pockets are large, other interactions, not present in Eq.~\eqref{eq:ansio_theory}, could become relevant and might change the low-energy physics. This is discussed on a qualitative level in Sec.~\ref{sec:conclusions}.

%%%%%%%%%%%%%%%%%%%%%%%%%%%%%%%%%%%%
\subsection{Three-pocket model}
%%%%%%%%%%%%%%%%%%%%%%%%%%%%%%%%%%%%

The isotropic Kitaev honeycomb model in a $[111]$ magnetic field features a $C_{3}^*$ symmetry of $120^\circ$ spin rotation about the field axis, combined with a $120^\circ$ lattice rotation about one site of the honeycomb lattice~\cite{janssen19}. 
A minimal Fermi-pocket model consistent with this symmetry and the half-filling constraint, Eq.~\eqref{eq:constraint}, comprises at least three Fermi pockets, such as, e.g., a particle-like pocket around the center $\boldsymbol\Gamma$ of the Brillouin zone and two hole-like pockets around its corners $\mathbf K$ and $\mathbf K'$.
Such a model was indeed suggested earlier as a candidate effective theory for the intermediate state observed in numerical studies of the antiferromagnetic Kitaev model in a $[111]$ magnetic field~\cite{jiang18, jiang19}.
It is given by the continuum action $S_3 = \int\rmd^2\mathbf x\rmd \tau \mathcal L_3$ with
\begin{equation}
\begin{split}
\mathcal{L}_{3}& = \psi_0^\dagger \left[\partial_\tau-\rmi ea_\tau-\frac{2m_1}{m_0}\tilde{\mu}+\frac{(-\rmi\boldsymbol\nabla-g\mathbf{a})^2}{2m_0}\right]\psi_0\pd\\
&\quad
+\psi_1^\dagger \left[\partial_\tau-\rmi ea_\tau+\tilde{\mu}-\frac{(-\rmi\boldsymbol\nabla-g\mathbf{a})^2}{2m_1}\right]\psi_1\pd
\\&\quad
+\psi_2^\dagger \left[\partial_\tau-\rmi ea_\tau+\tilde{\mu}-\frac{(-\rmi\boldsymbol\nabla-g\mathbf{a})^2}{2m_1}\right]\psi_2\pd\\
&\quad
+\frac{1}{4}f_{\mu\nu}f^{\mu\nu}
+u\left(\psi_0^\dagger\psi_0\pd\psi_1^\dagger\psi_1\pd+\psi_0^\dagger\psi_0\pd\psi_2^\dagger\psi_2\pd\right)\\
&\quad
+\tilde{u}\,\psi_1^\dagger\psi_1\pd\psi_2^\dagger\psi_2\pd,
\end{split}
\label{eq:iso_theory}
\end{equation}
where $\psi_0$ again describes spinon excitations near the particle-like pocket around the $\boldsymbol\Gamma$ point, while $\psi_1$ and $\psi_2$ describe spinon excitations near the hole-like pockets around the $\mathbf{K}$ and $\mathbf{K'}$ points, respectively.
Note that $\psi_1$ and $\psi_2$ are connected by inversion symmetry on the honeycomb lattice, but $\psi_0$ and $\psi_{1,2}$ are not symmetry related.
As a consequence, in addition to the particle-hole interpocket interaction parametrized by $u$, an independent hole-hole interpocket interaction parametrized by the coupling $\tilde{u}$ is allowed by symmetry and included in Eq.~\eqref{eq:iso_theory}.
Importantly, $\tilde u > 0$ ($\tilde u < 0$) corresponds to attraction (repulsion) between equally-charged holes.
As in Eq.~\eqref{eq:ansio_theory}, the parameter $\tilde{\mu}$ allows one to tune the sizes of the Fermi pockets, with the prefactor $2m_1/m_0$ in the first line of Eq.~\eqref{eq:iso_theory} again chosen to comply with Luttinger's theorem for half filling.
The $120^\circ$ rotational symmetry on the honeycomb lattice dictates that the effective band masses $m_1$ and $m_0$ are isotropic at the $\boldsymbol\Gamma$ and $\mathbf K$ points, respectively.
In the proposed realization in the antiferromagnetic Kitaev model in a $[111]$ magnetic field~\cite{jiang18, jiang19}, $\tilde{\mu}$ can be understood as a tuning parameter for the Lifshitz transition at $h_\mathrm{c}$, with $\tilde{\mu} \propto h_\mathrm{c} - h$.
For $\tilde{\mu} > 0$, the particle-like band corresponding to $\psi_0$ and the two hole-like bands corresponding to $\psi_1$ and $\psi_2$ cross the Fermi level in circles centered around $\boldsymbol\Gamma$ and  $\mathbf K$, $\mathbf K'$, respectively. This corresponds to the proposed U(1) spin liquid state for $h<h_\mathrm{c}$, and is illustrated in Figs.~\ref{fig:bandstructure}(a) and \ref{fig:fermi-pockets}(a).
For $\tilde{\mu} = 0$, the Fermi pockets shrink to isolated Fermi points at $\boldsymbol\Gamma$ and $\mathbf K$, $\mathbf K'$, corresponding to the proposed Lifshitz transition at $h=h_\mathrm{c}$, see Figs.~\ref{fig:bandstructure}(b) and \ref{fig:fermi-pockets}(b).
For $\tilde{\mu} < 0$, the spinon spectrum is fully gapped, Fig.~\ref{fig:bandstructure}(c). As monopoles of the compact U(1) gauge field proliferate in this situation~\cite{polyakov75, polyakov77}, the ground state in this case is conventional and expected to be adiabatically connected to the high-field polarized state~\cite{hickey19}.

In what follows, we again focus on the regime of small nonnegative $\tilde{\mu}$, in which a possible noncompactness of the gauge field is expected to be irrelevant as long as the spinon spectrum remains gapless~\cite{lee08}, and the finite extents of the Fermi pockets can be neglected. The scaling dimensions of the gauge couplings $e$ and $g$ and the particle-hole interpocket coupling $u$ are then the same as in the two-pocket model, Eq.~\eqref{eq:scaling-dims}, and similarly the hole-hole interpocket coupling $\tilde u$ has dimension $[\tilde u] = 2-d$.
Again, this allows a controlled perturbative RG analysis.

%%%%%%%%%%%%%%%%%%%%%%%%%%%%%%%%%%%%
\section{Renormalization group flow}
\label{sec:RG}
%%%%%%%%%%%%%%%%%%%%%%%%%%%%%%%%%%%%

\begin{figure*}
\includegraphics[width=\linewidth]{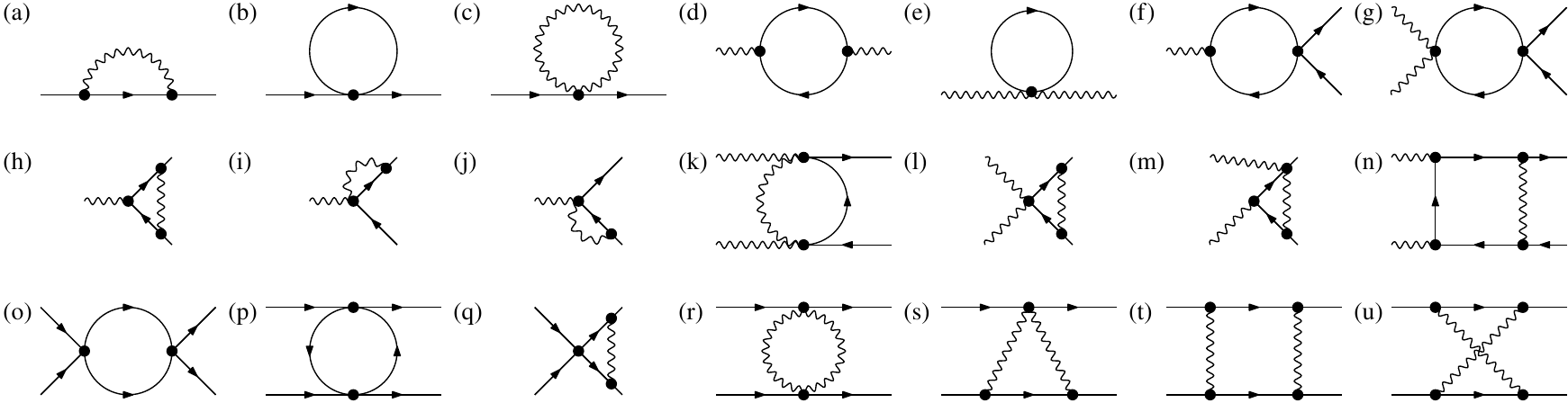}
\caption{Diagrammatic representation of fermion~(a)--(c) and gauge-field~(d)--(e) self-energies, as well as vertex corrections for gauge~(f)--(n) and interpocket~(o)--(u) couplings, at the one-loop order.
Solid and wavy inner lines denote fermion and gauge-field propagators, respectively.
Diagrams~(b) and (c) are independent of external momentum and therefore yield no contribution to the anomalous dimensions.
Diagrams~(d)--(g) involve closed particle-hole loops, and thus vanish identically in our models.
The gauge-field anomalous dimensions are therefore identically zero at the one-loop order.
Diagrams~(h)--(n) yield finite contributions that are related by means of Ward identities to the fermion anomalous dimensions arising from~(a).
In the two-pocket model, the particle-particle scattering diagram~(o) vanishes.
In the three-pocket model, it contributes only to the flow of $\tilde u$, while the particle-hole scattering diagram~(p) contributes only to the flow of $u$. The flows of $u$ and $\tilde u$ therefore do not mix.
Panels~(q) and (s) show representative diagrams for classes of six and two, respectively, triangle diagrams that contribute to the flows of the interpocket interactions.}
\label{fig:diagrams}
\end{figure*}

In this section, we perform a momentum-shell RG analysis at one-loop order, allowing us to study the fate of the U(1) spin liquid state with nested spinon Fermi pockets at low temperatures.

%%%%%%%%%%%%%%%%%%%%%%%%%%%%%%%%%%%%
\subsection{Two-pocket model}
%%%%%%%%%%%%%%%%%%%%%%%%%%%%%%%%%%%%

\begin{figure*}
\includegraphics[width=\linewidth]{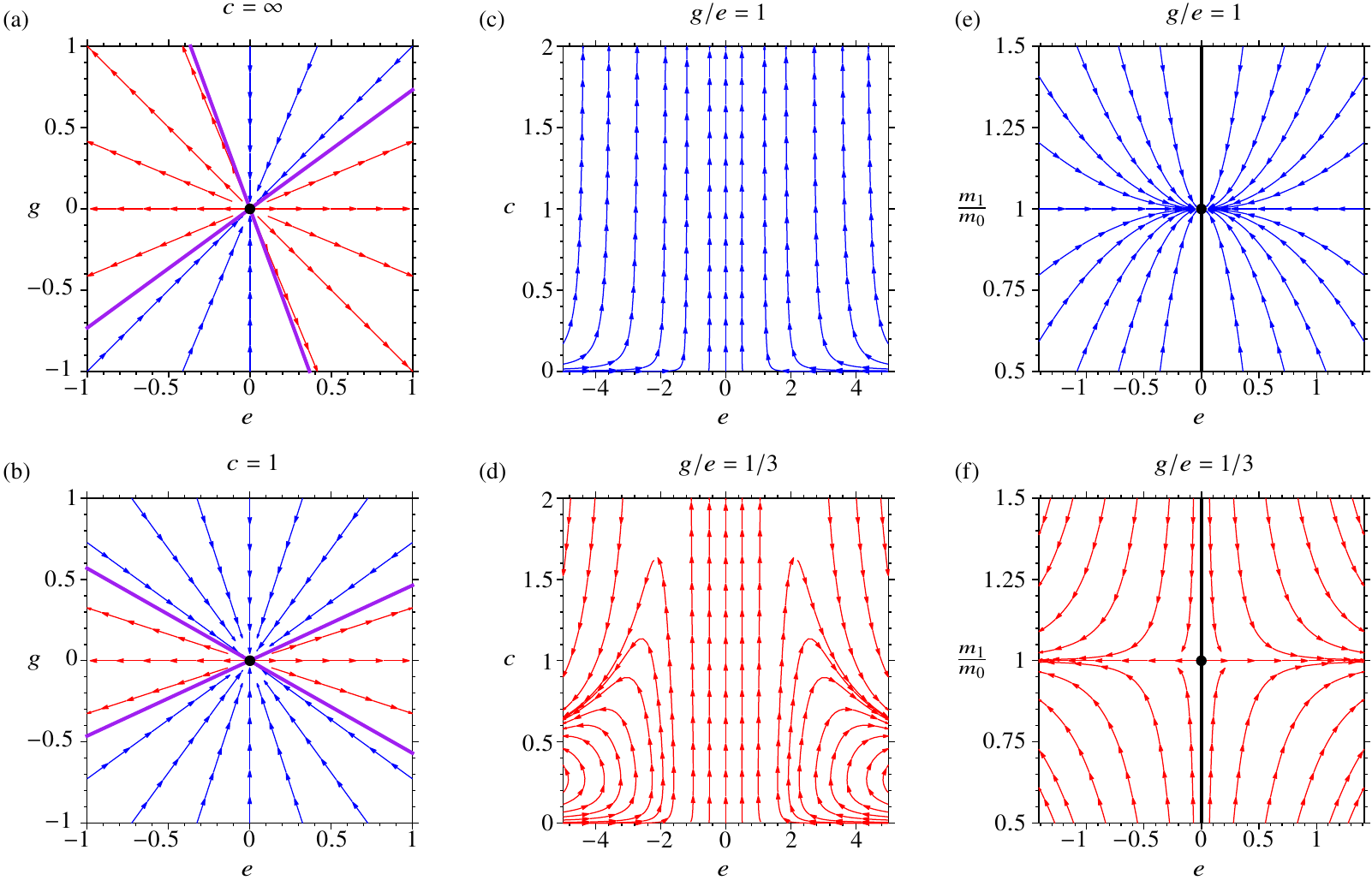}
\caption{RG flow of gauge couplings $e$ and $g$, speed of light $c$, and mass ratio $m_1/m_0$, for both the two-pocket and three-pocket models. Arrows denote flow towards infrared.
(a) Flow diagram in $e$-$g$ plane for fixed $c = \infty$. The separatrices (purple lines) divide the plane into distinct sectors, for which both $e$ and $g$ are relevant (red arrows) and irrelevant (blue arrows), respectively. The Gaussian fixed point $e = g = 0$ is marked as black dot.
(b) Same as (a), but for fixed $c = 1$, showing that the slope of the separatrices change as function of $c>0$, but the qualitative behavior for $(g/e)^2 \ll 1$ and $(g/e) \gg 1$ remains unchanged.
(c) Flow diagram in $e$-$c$ plane for fixed ratio $g/e = 1$ in the gauge-coupling-irrelevant sector [blue sector in (a),(b)], showing that $c$ flows to infinity in this sector.
(d) Same as (c), but for $g/e = 1/3$ in the gauge-coupling-relevant sector [red sector in (a),(b)], showing that $c$ initially increases, but eventually flows to a finite value $c_\star \in (0,\infty)$ in the deep infrared.
(e) Flow diagram in $e$-$\frac{m_1}{m_0}$ plane for fixed ratio $g/e = 1$ in the gauge-coupling-irrelevant sector [blue sector in (a),(b)], showing that the mass imbalance is irrelevant, in the sense that $m_1/m_0$ flows to one.
(f) Same as (e), but for $g/e = 1/3$ in the gauge-coupling-relevant sector [red sector in (a,b)], showing that the mass imbalance is also irrelevant in this sector.}
\label{fig:rgflow-e-g-c-m}
\end{figure*}

We start by discussing the two-pocket model defined in Eq.~\eqref{eq:ansio_theory} for vanishing renormalized $\tilde{\mu}=0$. We integrate out the fast modes with momenta in the shell $p \in (\Lambda/b,\Lambda)$, where $\Lambda$ is the ultraviolet cutoff, and all frequencies $\omega \in (-\infty,\infty)$. Evaluating the self-energy and vertex diagrams depicted in Fig.~\ref{fig:diagrams}(a)--(n) in Landau gauge leads to the flow equations for mass ratio $m_1/m_0$, speed of light $c$, and gauge couplings $e$ and $g$ as
\begin{align}
\frac{\rmd e}{\rmd\ln b}&=\frac{1}{2}F(g/e,c) e^3,
\label{eq:RGflow_aniso_a}\displaybreak[1]\\
\frac{\rmd g}{\rmd\ln b}&=\frac{1}{2}F(g/e,c) e^2 g,
\label{eq:RGflow_aniso_b}\displaybreak[1]\\
\frac{\rmd c}{\rmd\ln b}&=[1-F(g/e,c) e^2] c,
\label{eq:RGflow_aniso_c}\displaybreak[1]\\
\frac{\rmd (\tfrac{m_1}{m_0})}{\rmd\ln b}&=
[F(g/e,c) - F(g/e,c m_1/m_0)]e^2\tfrac{m_1}{m_0},
\label{eq:RGflow_aniso_d}
\end{align}
where we have rescaled $(e, g) \mapsto \sqrt{2\pi m_0}(e, g)$ and $c \mapsto \Lambda c /m_0$ in fixed $d=2$ spatial dimensions.
In the above flow equations, $F(g/e,c)$ is a second-order polynomial in the gauge coupling ratio $g/e$ and a rational function in the speed of light $c$.
For large $c \gg 1$, we have $F(g/e,c) = 1 -g/e -\frac12 (g/e)^2  + \mathcal O(1/c)$. Similarly, for general finite $c > 0$, we have $F(g/e,c) > 0$ [$F(g/e,c) < 0$] for $(g/e)^2 \ll 1$ [$(g/e)^2 \gg 1$].
Details of the renormalization procedure and a full expression of the function $F$ are given in the appendix.

Several comments on the above flow equations are in order:
(1)~The flow equations for $e$, $g$, $c$, and $m_1/m_0$ are independent of the particle-hole interpocket coupling $u$. While the renormalized values of these parameters are important for the flow of $u$, the latter does not couple back into the flow of the former. This is analogous to the situation in $(2+1)$-dimensional quantum electrodynamics with short-range interactions~\cite{kubota01, kaveh05, braun14, janssen16a, herbut16, gukov17}. We can therefore discuss the flow of $e$, $g$, $c$, and $m_1/m_0$ first and postpone the discussion of the flow of $u$ until later.
(2)~Similarly, the flow equations for $e$, $g$, and $c$ are independent of the mass ratio $m_1/m_0$. This can be understood as a consequence of the fact that diagrams involving closed particle-hole loops, such as those displayed in Figs.~\ref{fig:diagrams}(d)--(g), vanish in our models.
This is because the spinon band structures of our models do not allow particle-hole fluctuations of the same flavor.
(3)~At the present loop order, $e$ and $g$ have the same scaling dimension, such that their ratio $g/e$ is an RG invariant.  This again is a consequence of the vanishing of closed particle-hole loops, together with the Ward identity associated with the local U(1) symmetry.

As a first step, let us discuss the RG flow of the gauge couplings $e$ and $g$, together with the flow of $c$, all of which are independent of the mass ratio $m_1/m_0$.
Since $e$ and $g$ are marginal at tree level, their relevance or not upon the inclusion of loop fluctuations may depend on the initial values of the different parameters. In any case, however, $e$ and $g$ are either both relevant or both irrelevant, as their ratio remains constant.
Due to the structure of $F$, we find that $e$ and $g$ are irrelevant (relevant) for $(g/e)^2 \gg 1$ [$(g/e)^2 \ll 1$]. There are two critical ratios of $(g/e)^2$, for which $e$ and $g$ remain marginal. The values of these critical ratios depend on $c$.
Figures~\ref{fig:rgflow-e-g-c-m}(a) and~(b) show representative flow diagrams in the $e$-$g$ plane for two fixed values of $c$, illustrating the two gauge-coupling-irrelevant and gauge-coupling-relevant sectors, respectively, as well as the fact that the slope $g/e$ remains constant under the RG flow.
The speed of artificial light $c$ is a relevant parameter and initially increases under the RG flow. In the gauge-coupling-irrelevant sector for large ratios $(g/e) \gg 1$, it flows all the way up to infinity, while in the gauge-coupling-relevant sector for $(g/e)^2 \ll 1$ it converges to a finite infrared value $c_\star \in (0,\infty)$. The value of $c_\star$ again depends on the ratio $g/e$ and satisfies $F(g/e,c_\star) = 0$.
This is illustrated in Figs.~\ref{fig:rgflow-e-g-c-m}(c) and~(d), which show the RG flow projected onto the $e$-$c$ plane for two representative fixed ratios $g/e$ in the gauge-coupling-irrelevant and gauge-coupling-relevant sectors, respectively.
The fact that the speed of light's infrared value $c_\star$ is finite in the gauge-coupling-relevant sector near $(g/e)^2 \ll 1$ has a crucial consequence: As can be seen from Eq.~\eqref{eq:RGflow_aniso_c}, a finite $c_\star$ requires that $F(g/e,c_\star) e^2 \to 1$ in the infrared limit. In this limit, the flows of $e$ and $g$ [Eqs.~\eqref{eq:RGflow_aniso_a} and \eqref{eq:RGflow_aniso_b}] then effectively become $\rmd (e,g)/(\rmd \ln b) \simeq \frac12 (e,g)$, which means that $e$ and $g$ remain finite at all finite RG times. 
There is therefore no divergence of $e$ and $g$ any finite RG time in neither the gauge-coupling-irrelevant nor the gauge-coupling-relevant sector.

How does the flow of $e$, $g$, and $c$ affect the mass ratio $m_1/m_0$? As readily observable in Eq.~\eqref{eq:RGflow_aniso_d}, the flow of $m_1/m_0$ vanishes when the particle and hole bands have equal effective masses, $(m_1/m_0)_\star = 1$. We find that this equal-mass subspace is stable in the sense that $m_1/m_0$ flows to one for all initial finite values of $e$, $g$, and $c$.
This is illustrated in Figs.~\ref{fig:rgflow-e-g-c-m}(e) and~(f), which show the RG flow projected onto the $e$-$\frac{m_1}{m_0}$ plane again for the two representative fixed ratios $g/e$ in the gauge-coupling-irrelevant and gauge-coupling-relevant sectors, respectively.
In order to investigate the stability of the U(1) spin liquid state in the following, we may therefore focus on the equal-mass subspace and set the mass ratio to its stable infrared value $(m_1/m_0)_\star = 1$.

Let us now discuss how the gauge couplings $e$ and $g$ and the speed of artifical light $c$ influence the flow of the particle-hole interpocket interaction parametrized by $u$.
Evaluating the diagrams shown in Fig.~\ref{fig:diagrams}(o)--(u) for fixed $m_1/m_0 = 1$ leads to the flow equation
\begin{align}
    \frac{\rmd u}{\rmd\ln b}&=u^2 + G(g/e,c)e^2 u + H(g/e,c) e^4,
\label{eq:RGflow_aniso_e}
\end{align}
where we have rescaled $u \mapsto 2\pi u / m_0$, and $G(g/e,c)$ and $H(g/e,c)$ are second-order and fourth-order, respectively, polynomials in $g/e$ and rational functions in $c$.
For large $c \gg 1$, they satisfy $G(g/e,c) = 2 c^2 - 2 + 2 g/e + \frac12 (g/e)^2 +\mathcal O(1/c)$ and $H(g/e,c) = c^4 + (-1 + g/e) c^2 + \mathcal O(c)$. Similarly, for general finite $c > 0$, we have $G(g/e,c) > 0$ and $H(g/e,c)>0$ in both limits $(g/e)^2 \gg 1$ and $(g/e)^2 \ll 1$. Full forms of $G(g/e,c)$ and $H(g/e,c)$ are given in the appendix.
We note that the $G$ term in Eq.~\eqref{eq:RGflow_aniso_e} arises from the triangle diagram in Fig.~\ref{fig:diagrams}(q) and is therefore quadratic in $e$ and $g$ and linear in $u$, while the $H$ term results from the diagrams in Figs.~\ref{fig:diagrams}(r)--(u) and is quartic in $e$ and $g$, but independent of $u$.
These latter diagrams therefore generate a finite particle-hole interpocket coupling even in the absence of an initial $u$ at the microscopic scale. Importantly, the contribution is positive, leading to a repulsive interpocket interaction between equally-charged spinon excitations (or, equivalently, an attractive interpocket interaction between oppositely-charged excitations), in qualitative agreement with the situation of a simply-connected spinon Fermi surface~\cite{metlitski15}. This has important consequences for the stability of the U(1) spin liquid state with spinon Fermi pockets, as we see now.

Consider first the limit $e = g = 0$, which corresponds to the Gaussian fixed point in the gauge sector. In this limit, the flow of $u$ is simply $\frac{\rmd u}{\rmd \ln b} = u^2$, which has a unique fixed point for $u_\star = 0$. The interpocket coupling $u$ is marginally relevant (irrelevant) for $u>0$ ($u<0$). In fact, starting the flow for arbitrarily small $u>0$, we find that $u$ diverges at a finite RG time. This runaway flow corresponds to a divergent susceptibility in the channel associated with the particle-hole interpocket scattering and is to be understood as the onset of spontaneous symmetry breaking~\cite{chubukov12, fernandes17}. Starting the flow for $u<0$, however, we find that $u$ flows towards the Gaussian fixed point, corresponding to a stable Fermi-liquid-like phase.
We note that there is no Cooper pairing instability for attractive interactions~\cite{metlitski15} as a consequence of the nondegeneracy of the spinon bands and the fact that we focus on infinitesimal Fermi-pocket sizes.
For $e = g = 0$, the stability of nested Fermi pockets in our simple model hence depends on the initial sign of $u$.

This situation drastically changes upon the inclusion of gauge fluctuations for $e \neq 0$ or $g \neq 0$. As a finite gauge coupling induces a repulsive interpocket interaction $u>0$, the Gaussian fixed point becomes unstable in the presence of the U(1) gauge field. Integrating out numerically the RG flow in the full parameter space spanned by the $e$, $g$, $c$, and $u$, we \emph{always} find, for arbitrary initial values of $u$, a divergence of the interpocket coupling at finite RG time.
This divergence occurs not only in the gauge-coupling-relevant regime, but also in the gauge-coupling-irrelevant regime, in which $e$ and $g$ flow marginally to zero.
The latter can be understood analytically with the help of Eq.~\eqref{eq:RGflow_aniso_e} as follows: As both $G(g/e,c)$ and $H(g/e,c)$ are positive for all $g/e$ and $c>0$, any real zeros of the flow of $u$ for fixed $e$ and $c$ can only occur for negative~$u$. For large $c \gg 1$, these pseudo fixed points are located at $u_\star \simeq -e^2 c^2 + \mathcal O(c)$. As $c$ flows faster to infinity than $e$ to zero in the gauge-coupling-irrelevant sector, any pseudo fixed point is shifted towards $u_\star \to -\infty$ during the flow of $e$ and $c$, leaving behind only the runaway flow towards positive infinity.
Importantly, the divergence of the flow towards $u \to \infty$ for all initial values of $u$ and nonzero $e$ and/or $g$ occurs at finite RG time, and should hence be associated with the onset of spontaneous symmetry breaking~\cite{chubukov12, fernandes17}.
In the full parameter space spanned by $e$, $g$, $c$, $m_1/m_0$, and $u$, the U(1) spin liquid state is therefore ultimately unstable at low energy.
Similar gauge-coupling-driven instabilities have been discussed in a number of relativistic~\cite{%
%QED3:
kubota01, kaveh05, braun14, janssen16a, herbut16, gukov17, wang17,
%
%QCD4:
gies06, braun11,
%
%AdS/CFT:
faedo20,
%
%Abelian Higgs:
halperin74, ihrig19, nogueira19}
and nonrelativistic~\cite{%
herbut14, janssen17a}
gauge theories with matter content in two and three spatial dimensions. We note, however, in contrast to these previous works, the instability here, at least for the gauge-coupling-irrelevant case, does not occur as a consequence of a fixed-point annihilation mechanism, but rather should be understood to originate from a running off of the pseudo fixed points to infinity negative coupling~\cite{kaplan09}.
Properties of the resulting new ground state will be discussed in Secs.~\ref{sec:MFT} and~\ref{sec:discussion}.

%%%%%%%%%%%%%%%%%%%%%%%%%%%%%%%%%%%%
\subsection{Three-pocket model}
%%%%%%%%%%%%%%%%%%%%%%%%%%%%%%%%%%%%

\begin{figure*}
\includegraphics[width=\linewidth]{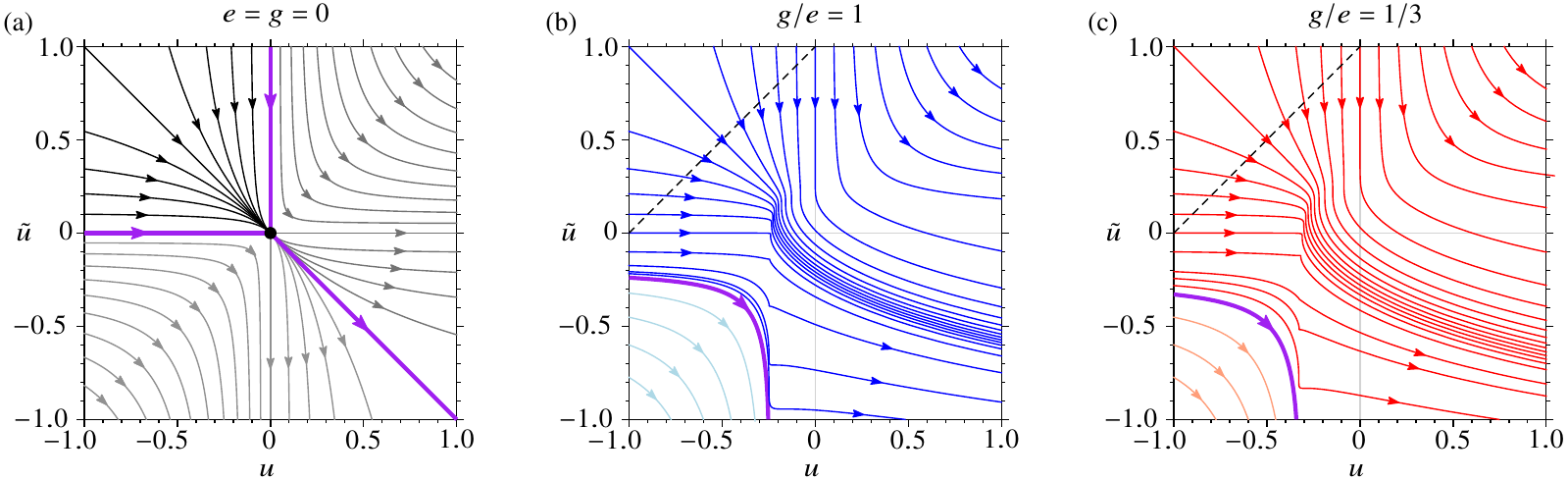}
\caption{RG flow of interpocket interactions $u$ and $\tilde u$ in the three-pocket model for $m_1/m_0 = 1$. Arrows denote flow towards infrared.
(a) Flow diagram in $u$-$\tilde{u}$ plane for fixed $e=g=0$, illustrating the Gaussian fixed point at $(u_\star, \tilde u_\star) = (0,0)$, and the division of parameter space into three distinct regions: For $u<0$ and $\tilde u>0$, both couplings are marginally irrelevant. For $u>0$ and $\tilde u > -u$, there is a runaway flow towards $u \to \infty$ and $\tilde u/u \to 0$, while for $\tilde u < 0$ and $u < - \tilde u$, there is a runaway flow towards $\tilde u \to -\infty$ and $u/\tilde u \to 0$.
(b) Integrated flow in full parameter space spanned by $(e,g,c,u,\tilde u)$ and projected onto the $u$-$\tilde u$ plane, in the gauge-coupling-irrelevant sector [blue sector in Fig.~\ref{fig:rgflow-e-g-c-m}(a),(b)]. Here, we have chosen fixed initial values of $(e,g,c) = (0.01, 0.01,1)$ and different initial values of $(u,\tilde u)$ along the dashed line depicted in the figure. Despite the marginal irrelevance of $e$ and $g$, there is a runaway flow for all initial values of $u$ and $\tilde u$. The thick purple curve separates the upper right subsector, in which $u \to \infty$ and $\tilde u / u \to 0$ (heavy blue), from the lower left subsector, in which $\tilde u \to -\infty$ and $u / \tilde u \to 0$ in the infrared limit (light blue).
(c) Same as (b), but in the gauge-coupling-relevant sector [red sector in Fig.~\ref{fig:rgflow-e-g-c-m}(a),(b)] for fixed initial values of $(e,g,c) = (0.03, 0.01,1)$. Again, there is a runaway flow for all initial values of $u$ and $\tilde u$, with leading divergence $u \to +\infty$ (heavy red) or $\tilde u \to -\infty$ (light red).}
\label{fig:rgflow-u-utilde}
\end{figure*}

In the previous subsection, we have seen that the U(1) spin liquid state with two perfectly nested spinon Fermi pockets is ultimately unstable at low temperatures due to a divergence of the particle-hole interpocket interaction at finite RG time. We now show that a similar instability also occurs in the three-pocket model defined in Eq.~\eqref{eq:iso_theory} with two independent interpocket interactions, parametrized by $u$ and $\tilde u$.

To begin with, we note that the flow equations for $e$, $g$, $c$, and $m_1/m_0$ in the three-pocket model are identical to those of the two-pocket model, Eqs.~\eqref{eq:RGflow_aniso_a}--\eqref{eq:RGflow_aniso_d}. 
This is another consequence of the fact that Feynman diagrams with closed particle-hole loops vanish in the present type of models.
Hence, the previous discussion of the flows of $e$, $g$, $c$, and $m_1/m_0$, as well as the flow diagrams depicted in Fig.~\ref{fig:rgflow-e-g-c-m}, fully apply also to the present three-pocket model.
The roles of the two different interpocket interactions parametrized by $u$ and $\tilde u$ remain to be examined. Evaluating the corresponding diagrams, shown in Figs.~\ref{fig:diagrams}(o)--(u), for the three-pocket model, leads to the flow equations
\begin{align}
\frac{\rmd u}{\rmd\ln b}&=u^2 + G(g/e,c)e^2 u + H(g/e,c) e^4,
\label{eq:RGflow_e}\displaybreak[1]\\
\frac{\rmd \tilde{u}}{\rmd\ln b}&=-\tilde{u}^2 - \tilde{G}(g/e,c)e^2\tilde{u} - H(g/e,c) e^4.
\label{eq:RGflow_f}
\end{align}
where we have rescaled $(u,\tilde u) \mapsto 2\pi (u,\tilde u)/m_0$. The functions $G(g/e,c)$ and $H(g/e,c)$ are the same as those in Eq.~\eqref{eq:RGflow_aniso_e}, while $\tilde G(g/e,c)$ is another second-order polynomial in $g/e$ and rational function in $c$, with $\tilde G(g/e,c) = 2 c^2 - \frac12 (g/e)^2 + \mathcal O(1/c)$ for $c \gg 1$. An explicit expression of $\tilde G(g/e,c)$ for general $c>0$ is given in the appendix.

The flow in the $u$-$\tilde u$ plane is depicted for different initial values of the gauge couplings $e$ and $g$ in Fig.~\ref{fig:rgflow-u-utilde}. For vanishing gauge couplings, $e=g=0$, there is a Gaussian fixed point at $(u_\star,\tilde u_\star)=(0,0)$, which attracts the flow in the region $u<0$ and $\tilde u>0$, see Fig.~\ref{fig:rgflow-u-utilde}(a). For $u > 0$ and $\tilde u>-u$, there is a runaway flow towards $u \to +\infty$ and $\tilde u/ u \to 0$. In the remaining region for $\tilde u<0$ and $u<-\tilde u$, on the other hand, there is a runaway flow towards $\tilde u \to -\infty$ and $u/\tilde u \to 0$.
Figure~\ref{fig:rgflow-u-utilde}(b) shows the numerically integrated flow in the five-dimensional parameter space spanned by $e$, $g$, $c$, $u$, and $\tilde u$, projected onto the $u$-$\tilde u$ plane, using initial values for $e$ and $g$ in the gauge-coupling-irrelevant sector.
As in the two-pocket model, there is no fixed point remaining at any finite $u$ and $\tilde u$, destabilizing the region of $u<0$ and $\tilde u >0$, and leaving behind the runaway flow towards $u \to \infty$.
We have explicitly verified this picture by integrating out the flow for various sets of initial parameter values, confirming that the runaway flow occurs for arbitrary initial values of $u$ and $\tilde u$ always at finite RG times, even when $e$ and $g$ flow to zero in the gauge-coupling-irrelevant sector.
In analogy to the two-pocket model, we interpret this destabilization to arise from a running-off of the pseudo fixed points towards $u_\star \to -\infty$, and it signals the onset of spontaneous symmetry breaking.
A similar destabilization occurs for initial values of $e$ and $g$ in the gauge-coupling-relevant sector, as depicted in Fig.~\ref{fig:rgflow-u-utilde}(c). Again, the fact that the gauge couplings are finite implies the absence of any fixed point at finite couplings in the $u$-$\tilde u$~plane and the divergence of an interpocket coupling at finite RG time.

%%%%%%%%%%%%%%%%%%%%%%%%%%%%%%%%%%%%
\section{Spinon gap opening}
\label{sec:MFT}
%%%%%%%%%%%%%%%%%%%%%%%%%%%%%%%%%%%%

The RG analysis discussed in the previous section shows that the U(1) spin liquid state with nested Fermi pockets is unstable due to a divergent interpocket coupling at finite RG times.
There is a corresponding nesting vector associated with each interpocket interaction, which depends on the location of the Fermi pockets within the Brillouin zone in the particular microscopic realization of the model.
In the proposed realization of the two-pocket model for the quadruple point in the phase diagram of the anisotropic Kitaev model in a $[111]$ magnetic field~\cite{jiang18}, the coupling $u$ connects states separated by the nesting vector $\mathbf Q_\text{nesting} = \mathbf M$. In this realization, a divergent $u$ therefore corresponds to a divergent susceptibility at wavevector $\mathbf M$ and appears to favor a long-range-ordered ground state with ordering wavevector $\mathbf M$. Such a state doubles the real-space unit-cell size in the direction of $\mathbf M$, and hence breaks part of the lattice translational symmetry spontaneously.
In the proposed realization of the three-pocket model for the field-induced intermediate phase of the isotropic Kitaev model~\cite{jiang18, jiang19}, the coupling $u$ connects states in the particle-like pocket at $\boldsymbol{\Gamma}$ with states in the hole-like pockets at $\mathbf K$ and $\mathbf K'$, while the coupling $\tilde u$ connects states in the two different hole-like pockets. In both channels, the corresponding nesting vector is $\mathbf Q_\text{nesting} = \mathbf K$, see Fig.~\ref{fig:fermi-pockets}.
A divergence of $u$ and/or $\tilde u$ hence suggests a long-range-ordered ground state with ordering wavevector $\mathbf K$ in this case, which triples the unit-cell size, but preserves the $120^\circ$ rotational symmetry on the honeycomb lattice~\cite{janssen16b, chern17}.
This illustrates that the precise nature of the resulting ground state arising from the divergence of the interpocket couplings is nonuniversal and depends on the particular microscopic realization of the effective continuum model.
Instead of a full characterization, in this section, we therefore content ourselves with pointing out some general universal features of the low-temperature phase, which can be inferred readily from our effective modeling. In particular, we will show that the divergence of the interpocket couplings leads to a full gap opening in the spinon band structure at low temperatures.
A further characterization of the gapped ground state, taking microscopic properties of particular realizations of our effective models into account, is postponed to Sec.~\ref{sec:discussion}.

\begin{table*}[tb]
\caption{Field content, Hubbard-Stratonovich decoupling channel, and spinon gap in the two-pocket model with particle-hole interpocket coupling $u>0$ (first row), and in the three-pocket model for dominant particle-hole interpocket coupling $u>0$ (second row) and dominant hole-hole interpocket coupling $\tilde u < 0$ (third row), respectively. While the particle-hole interaction in the two-pocket model and the hole-hole interaction in the three-pocket model can be decoupled with a single-component real order-parameter field $\Phi = \varphi$ and $\Phi = \phi$, respectively, the particle-hole interaction in the three-pocket model requires a decoupling in terms of a two-component real order-parameter field $\Phi = (\chi_1, \chi_2)$. The mean-field ground state in each of the three cases is characterized by a full gap $\Delta > 0$ in the spinon spectrum for all finite $g = h^2/r>0$.}
\label{tab:decoupling_channels}
\begin{tabular*}{\linewidth}{@{\extracolsep{\fill} } l l l l l l}
\hline\hline
Model & Dominant coupling & Spinon fields & Order parameter & Decoupling channel & Spinon gap \\ 
& $g = h^2/r > 0$ & $\Psi$, $\Psi^\dagger$ complex & $\Phi$ real & $\Phi \sim \Psi^\dagger A \Psi$ & $\Delta \propto h\langle \Phi \rangle$
\\ \hline
%
% $\langle \Phi \rangle  = - \frac{h}{r} \langle\Psi^\dagger A \Psi\rangle$
%
Two-pocket &
$g = u > 0$ & 
$\Psi = \begin{pmatrix}\psi_0\\ \psi_1\end{pmatrix}$ &
$\Phi = \varphi$ & 
$A=\begin{pmatrix}1&0\\0&-1\end{pmatrix}$ &
$\Delta = 2h\langle \varphi \rangle = \frac{\Lambda^2}{2\pi} u > 0$
\\
Three-pocket & 
$g = u > 0$ & 
$\Psi = \begin{pmatrix}\psi_0\\ \psi_1\\\psi_2\end{pmatrix}$ & 
$\Phi = \begin{pmatrix}\chi_1\\ \chi_2\end{pmatrix}$ &
$A=\begin{pmatrix}
\left(\begin{smallmatrix}1&0&0\\0&-1&0\\0&0&0\end{smallmatrix}\right) \vspace{.5ex} \\
\left(\begin{smallmatrix}1&0&0\\0&0&0\\0&0&-1\end{smallmatrix}\right)
\end{pmatrix}$ &
$\Delta = 3h \langle \chi_{1} \rangle = 3h \langle \chi_{2} \rangle = \frac{3\Lambda^2}{4\pi} u > 0$
\\
&
$g = -\tilde{u} > 0$ & 
$\Psi = \begin{pmatrix}\psi_0\\ \psi_1\\\psi_2\end{pmatrix}$ & 
$\Phi = \phi$ & 
$A=\begin{pmatrix}0&0&0\\0&1&0\\0&0&1\end{pmatrix}$ &
$\Delta = - h \langle \phi \rangle = - \frac{\Lambda^2}{4\pi} \tilde u > 0$ 
\\
\hline\hline
\end{tabular*}
\end{table*}

To make progress analytically, we focus on the infrared regime in which the interpocket couplings are large and provide the dominant contribution to the partition function. While the fluctuations of the gauge field are crucial for the generation of repulsive interpocket couplings in the first place, they can be safely neglected in this low-energy regime.
This suggests a mean-field analysis of the Hubbard-Stratonovich-decoupled interaction channel corresponding to the interpocket coupling with the strongest RG divergence~\cite{wang14}.
In our models, the situation is simplified by the fact that there is, for each set of initial couplings, a unique dominant decoupling channel associated with the divergent RG flow.
In the two-pocket model, we have $u \to \infty$, associated with the channel $\varphi \sim \psi_0^\dagger \psi_0\pd - \psi_1^\dagger \psi_1\pd$. A finite vacuum expectation value $\langle\varphi\rangle \neq 0$ corresponds to a shift in energy of the particle-like and hole-like spinon bands relative to each other and, as we show explicitly below, leads to a gap opening in the spinon spectrum.
In the three-pocket model, there are two different regimes with either $u \to \infty$ and $\tilde u/u \to 0$ or $\tilde u \to -\infty$ and $u/\tilde u \to 0$, depending on the initial values of the couplings.
In the former regime, the associated decoupling channel is $(\chi_1,\chi_2) \sim (\psi_0^\dagger \psi_0\pd - \psi_1^\dagger\psi_1\pd, \psi_0^\dagger \psi_0\pd - \psi_2^\dagger\psi_2\pd)$, which again shifts the particle-like and hole-like bands relative to each other, preserving the symmetry between the two hole-like pockets if $\langle \chi_1 \rangle = \langle \chi_2 \rangle \neq 0$.
In the latter regime, the associated decoupling channel is $\phi \sim (\psi_1^\dagger \psi_1\pd + \psi_2^\dagger \psi_2\pd)$ and corresponds to a simultaneous shift of the two hole-like bands only.
Within our two models, there are therefore in total three cases to be distinguished. In all three cases, the corresponding Hubbard-Stratonovich transformation can be written on the level of the partition function $\mathcal Z$ generically as
\begin{align}
\mathcal{Z} & = \int D\Psi D\Psi^\dagger\rme^{-\left\{S_0[\Psi,\Psi^\dagger]-\frac{g}{2}(\Psi^\dagger A \Psi)^2\right\}}
\nonumber \\
& \propto \int D\Psi D\Psi^\dagger D\Phi\rme^{-\left\{S_0[\Psi,\Psi^\dagger]+\frac{r}{2} \Phi^2 + h \Phi \cdot (\Psi^\dagger A \Psi)\right\}},
\label{eq:HST}
\end{align}
where we have introduced the two-component (three-component) complex spinor fields $\Psi$ and $\Psi^\dagger$, with $\Psi = (\psi_0, \psi_1)^\top$ [$\Psi = (\psi_0,\psi_1,\psi_2)^\top$], and the real bosonic order-parameter field $\Phi$, with $\Phi = \varphi$ [$\Phi = (\chi_1, \chi_2)$ and $\Phi = \phi$, respectively] for the case(s) of the two-pocket (three-pocket) model.
The functional $S_0$ denotes the noninteracting Gaussian part of the fermionic action and the parameter $g>0$ represents the particular repulsive interpocket coupling, with $g = u$ in the two-pocket model and $g = u$ ($g = - \tilde u$) in the three-pocket model for dominant positive $u$ (dominant negative $\tilde u$).
The real and symmetric $2\times 2$ and $3\times 3$ matrices $A$, respectively, define the corresponding decoupling channel and are given explicitly for the three different cases in Table~\ref{tab:decoupling_channels}.
The Hubbard-Stratonovich transformation in Eq.~\eqref{eq:HST} becomes exact provided that we identify $g \equiv h^2/r$ for $r>0$. Without loss of generality, we further assume $h\geq 0$.
Then, positive expectation values $\langle\varphi\rangle > 0$ and $\langle \chi_{1,2}\rangle > 0$ correspond to an upward (downward) shift in energy of the particle-like (hole-like) bands for the cases of dominant particle-hole coupling $u > 0$, while negative $\langle \phi \rangle < 0$ corresponds to a downward shift of the hole-like bands for the case of dominant hole-hole coupling $\tilde u < 0$.

\begin{figure}[tb]
\includegraphics[width=\linewidth]{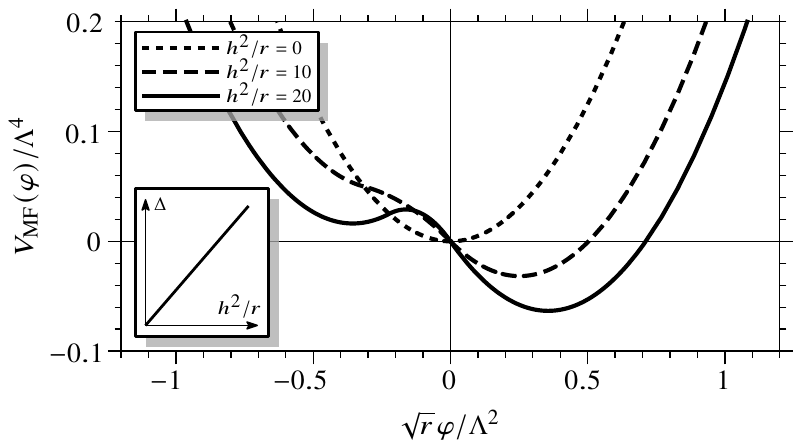}
\caption{Mean-field potential in two-pocket model as function of order parameter $\varphi$. For $h^2/r = 0$, the minimum is at $\varphi=0$, while it moves to positive $\varphi > 0$ for all finite $h^2/r>0$. This corresponds to a spinon gap $\Delta \propto h^2/r$, as schematically shown in the inset.}
\label{fig:potential}
\end{figure}

Since the Hubbard-Stratonovich-decoupled action is quadratic in $\Psi$ and $\Psi^\dagger$, the fermions can be integrated out, yielding an effective action for the order-parameter field $\Phi$. On the mean-field level, the expectation value of the order parameter is then obtained by minimizing the effective action with respect to $\Phi$, neglecting the fluctuations in $\Phi$. This approximation can be understood as the leading order of a controlled $1/N$ expansion, where $N$ is the number of spinon Fermi pockets in the model.
We demonstrate the calculation explicitly for the case of dominant positive $u$ in the two-pocket model, corresponding to the first row in Table~\ref{tab:decoupling_channels}.
For the three-pocket model, the calculation is analogous, and we will restrict the presentation to the discussion of the main results in this case.
In the two-pocket model, the mean-field effective potential reads
\begin{multline}
V_\text{MF}(\varphi) = 
\frac{r}{2}\varphi^2 
-\int_{-\infty}^\infty \frac{d\omega}{2\pi} \int^\Lambda \frac{d^2\mathbf{p}}{(2\pi)^2}\left[\ln(\rmi\omega+\mathbf{p}^2 + h \varphi)
\right. \\ \left.
+\ln(\rmi\omega-\mathbf{p}^2-h\varphi)\right],
\end{multline}
where we have reintroduced the ultraviolet cutoff $\Lambda$, and assumed units in which the effectice band masses are set to $m_0 \equiv 1/2$ and $m_1 \equiv 1/2$, exploiting again the property that the mass ratio $m_1/m_0 = 1$ in the infrared.
Up to a physically irrelevant constant $V_\text{MF}(0)$, the integral is convergent in the sense of the Cauchy principle value, and evaluates to 
\begin{equation}
V_\text{MF}(\varphi) =
\begin{cases}
\frac{r}{2}\varphi^2+\frac{\Lambda^2}{4\pi}h\varphi+\frac{\Lambda^4}{4\pi} 
& \mathrm{for}\,\varphi< - \frac{\Lambda^2}{h},\\
(\frac{r}{2}-\frac{h^2}{4\pi})\varphi^2-\frac{\Lambda^2}{4\pi}h\varphi
& \mathrm{for}\, - \frac{\Lambda^2}{h} < \varphi < 0,\\
\frac{r}{2}\varphi^2-\frac{\Lambda^2}{4\pi}h\varphi
& \mathrm{for}\,\varphi>0,
\end{cases}
\label{eq:V_eff}
\end{equation}
where we have set $V_\text{MF}(0) \equiv 0$ without loss of generality.
Note that $V_\text{MF}(\varphi)$ is continuously differentiable at all $\varphi$.
The effective potential is plotted for different values of $h^2/r$ in Fig.~\ref{fig:potential}.
The minimum of the effective potential $V_\text{MF}(\varphi)$ corresponds to the mean-field expectation value $\langle\varphi\rangle$.
In the noninteracting case for $h^2/r=0$, we have $\langle\varphi\rangle=0$.
For any finite $h^2/r > 0$, however, the minimum shifts towards positive $\sqrt{r}\langle\varphi\rangle=\frac{\Lambda^2}{4\pi}\sqrt{{h^2}/{r}}>0$, corresponding to a full gap $\Delta = 2h \langle \varphi \rangle = \frac{\Lambda^2}{2\pi}\frac{h^2}{r} > 0$ in the spinon spectrum.
The linear opening of the gap $\Delta$ as a function of $h^2/r$ is shown in the inset of Fig.~\ref{fig:potential}.
Similarly, in the case of the three-pocket model for dominant positive $u$, we find that the potential $V_\text{MF}(\chi_1, \chi_2)$ for the corresponding order parameter $\Phi = (\chi_1,\chi_2)$ becomes minimal for $\langle\chi_1\rangle = \langle \chi_2 \rangle > 0$, while for dominant negative $\tilde u$, we find that $V_\text{MF}(\phi)$ attains its minimum at $\langle \phi \rangle <0$.
These results imply that in all three cases, the RG divergence of the repulsive interpocket interactions correspond to a full gap opening in the spinon spectrum.
This conclusion is consistent with the general expectation that the mean-field energy is minimized when the spectral gap is maximized~\cite{janssen15, ray21}.
For each of the three cases, mean-field gap $\Delta$ and order-parameter expectation value $\langle \Phi \rangle$ are given explicitly as functions of the respective interpocket coupling in the last column of Table~\ref{tab:decoupling_channels}.

%%%%%%%%%%%%%%%%%%%%%%%%%%%%%%%%%%%%
\section{Low-temperature phase}
\label{sec:discussion}
%%%%%%%%%%%%%%%%%%%%%%%%%%%%%%%%%%%%

In the previous two sections, we have argued that a U(1) spin liquid state with nested spinon Fermi pockets is unstable at low temperatures towards a long-range-ordered ground state characterized by a full gap in the spinon spectrum.
In this section, we shall further characterize the nature of the resulting low-temperature phase, taking properties of proposed microscopic realizations of our continuum models into account.

An important characteristic of all proposed realizations of U(1) spin liquids with spinon Fermi pockets~\cite{jiang18, jiang19, zou18, hickey19, patel19, sodemann21} is the fact that the emergent gauge field in the lattice model is actually compact. This allows instanton events that change the magnetic flux within a plaquette of the lattice and create a monopole of the gauge field.
In the presence of a finite density of states at the Fermi level, monopole creation operators are RG irrelevant and our continuum field theory, which assumes a noncompact U(1) gauge field, is a valid low-energy description~\cite{lee08}.
However, once a gap in the spinon spectrum has opened up as a consequence of the RG divergence of an interpocket coupling, the noncompactness of the gauge field needs to be taken into account.
In fact, in the absence of any low-energy matter degrees of freedom, monopole creation operators become RG relevant and monopoles start to proliferate below a certain temperature scale~\cite{polyakov75, polyakov77}. This leads to confinement of spinons and a fully conventional ground state that features magnetic or nonmagnetic long-range order.

\begin{figure}[tb]
\includegraphics[width=\linewidth]{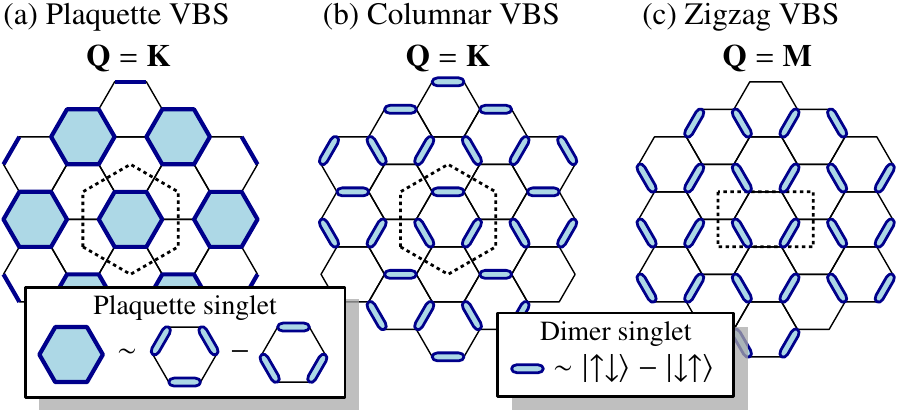}
\caption{Candidate quantum paramagnetic ground states of models for U(1) spin liquids with nested spinon Fermi pockets on the honeycomb lattice.
For the three-pocket model with spinon Fermi pockets around $\boldsymbol\Gamma$ and $\mathbf K$, $\mathbf K'$ points~\cite{jiang18, jiang19}, the low-temperature instability is towards a state with ordering wavevector $\mathbf Q = \mathbf K$, such as (a) the plaquette valence bond solid (VBS)~\cite{zhu13, ganesh13b, ganesh13a} or (b) the columnar VBS~\cite{read90, fouet01, albuquerque11, pujari13, pujari15}.
Both states feature six-site unit cells (dotted hexagons), with the columnar VBS being characterized by a fixed pattern of dimer singlets, while the plaquette VBS is characterized by a pattern of plaquette singlets, each of which consists of an antisymmetric combination of the Kekul\'e structure, as depicted in the insets.
The columnar VBS is also known as Kekul\'e VBS~\cite{hou07, ryu09, fu11, pujari15} or Read-Sachdev~\cite{read90, fouet01} state in the literature.
For the two-pocket~\cite{jiang18} and four-pocket~\cite{patel19} models with spinon Fermi pockets around $\boldsymbol\Gamma$ and $\mathbf M$ points, the divergent interpocket interaction can alternatively also drive an instability towards a state with ordering wavevector $\mathbf Q = \mathbf M$, such as (c) the zigzag VBS with four-site unit cell (dotted rectangle). In this state, neighboring spins along zigzag chains on the honeycomb lattice form dimer singlets. It breaks lattice rotational symmetry and can be understood as a four-sublattice version of the staggered VBS~\cite{mulder10, xu11}.}
\label{fig:vbs-states}
\end{figure}

As argued in the beginning of the previous section, the ordering wavevector of the state favored by the divergent interpocket coupling is given by the nesting vector $\mathbf Q_\text{nesting}$, which connects spinon excitations in different Fermi pockets of the spin liquid state.
For the proposed realization of the three-pocket model in the antiferromagnetic Kitaev model in a $[111]$ magnetic field~\cite{jiang18, jiang19}, we have $\mathbf Q_\text{nesting} = \mathbf K$, corresponding to the corners of the Brillouin zone, see Fig.~\ref{fig:fermi-pockets}.
The divergent interpocket coupling in this case hence favors a state with ordering wavevector $\mathbf Q = \mathbf K$, which breaks lattice translational symmetry but preserves the 120$^\circ$ lattice rotational symmetry. The corresponding hexagonal unit cell consists of six sites. There are a number of different magnetic~\cite{janssen16b, chern17} and nonmagnetic~\cite{read90, fouet01, albuquerque11} states on the honeycomb lattice known that feature this ordering wavevector. 
However, the fact that monopoles of the U(1) gauge field proliferate at low energy suggests that the ground state is paramagnetic~\cite{read90, xu10}.
In fact, on the honeycomb lattice, the monopole creation operator has precisely the same quantum numbers as the order parameter for a VBS order at $\mathbf Q = \mathbf K$~\cite{xu10, fu11, pujari13, pujari15}. 
The divergence of the interpocket coupling and the proliferation of monopoles therefore conspire to destabilize the U(1) spin liquid state with nested spinon Fermi pockets at the $\mathbf K$ points towards a quantum paramagnetic ground state with a six-site unit cell and VBS order.
This leaves us with two possible ground states, depicted in Figs.~\ref{fig:vbs-states}(a) and~(b). These states are known as plaquette VBS~\cite{fouet01, albuquerque11, zhu13, ganesh13b, ganesh13a} and columnar VBS~\cite{read90, pujari13, pujari15} in the literature.
In these quantum paramagnetic states, the fractionalized spinons are gapped and confined, but the low-energy spectrum may feature gapless spin-singlet excitations in the thermodynamic limit~\cite{albuquerque11}.

A similar analysis is possible for the situation with spinon Fermi pockets at the $\boldsymbol\Gamma$ and $\mathbf M$ points in the Brillouin zone, such as in the two-pocket and four-pocket models proposed in Refs.~\cite{jiang18} and \cite{patel19}, respectively.
The difference in this case, however, is that the divergent interpocket coupling favors a state with ordering wavevector $\mathbf Q = \mathbf M$, corresponding to a four-site rectangular unit cell.
This may be a quantum paramagnetic state, such as the zigzag VBS depicted in Fig.~\ref{fig:vbs-states}, which can be understood as a four-sublattice version of the staggered VBS state~\cite{mulder10, xu11}, but magnetically ordered states, such as stripy and zigzag antiferromagnets~\cite{fouet01, chaloupka10, chaloupka13}, are equally well possible.
In these models, the plaquette and columnar VBS states favored by the proliferating monopoles~\cite{xu10, fu11} therefore compete with such four-site-unit-cell states. The question which one will eventually win depends on microscopic parameters of the particular system at hand and is beyond the scope of the present effective modeling.

%%%%%%%%%%%%%%%%%%%%%%%%%%%%%%%%%%%%
\section{Conclusions}
\label{sec:conclusions}
%%%%%%%%%%%%%%%%%%%%%%%%%%%%%%%%%%%%

In this work, we have studied the fate of U(1) spin liquids with nested spinon Fermi pockets in two spatial dimensions.
We have argued that the emergent U(1) gauge field generically generates a repulsive interaction between equally-charged spinon excitations and drives a divergence of an interpocket coupling during the RG flow. This divergence can be understood as a nesting instability of the U(1) spin liquid state.
The instability induces a full gap in the spinon spectrum at low temperatures. This causes a proliferation of monopoles of the compact U(1) gauge field and leads to a conventional long-range-ordered ground state. In this state, all gauge-non-invariant excitations, such as the fractionalized spinons, are confined.
This main conclusion of our work has been shown explicitly within two minimal models, which feature two and three, respectively, perfectly nested spinon Fermi pockets in the Brillouin zone.
From this analysis, however, it is clear that a similar instability should be expected also in models with larger numbers of spinon Fermi pockets.
An effective description of the four-pocket spin liquid state suggested in Ref.~\cite{patel19}, for instance, would consist of three symmetry-related spinon fields $\psi_i$, $i=1,2,3$, corresponding to the pockets near the three $\mathbf M$ points in the hexagonal Brillouin zone, in addition to the spinon field $\psi_0$ that corresponds to the pocket near the $\boldsymbol\Gamma$ point. This again allows only two interpocket interaction channels
\begin{align}
u \sum_{i} (\psi_0^\dagger\psi\pd_0 \psi_i^\dagger\psi\pd_i)
+    
\tilde u \sum_{i<j} (\psi_i^\dagger\psi\pd_i \psi_j^\dagger\psi\pd_j),
\end{align}
in close analogy to the three-pocket model of Eq.~\eqref{eq:iso_theory}.
A similar divergence of an interpocket coupling should therefore occur also in this case.
The instability of U(1) spin liquids with nested spinon Fermi pockets can also be understood in terms of a spin-Peierls transition, as an example of the general tendency of frustrated low-dimensional spin systems to form VBS order~\cite{read90}. A similar spin-Peierls instability has recently been studied in the context of three-dimensional $\mathbbm Z_2$ spin liquids with Majorana Fermi surfaces, although, in this case, the result of the instability is not a VBS state, but another spin liquid~\cite{hermanns15}.

For our RG calculations, we have focused on the limit of infinitesimally small spinon Fermi pockets.
We argued that monopole operators are likely irrelevant, as long as the spinon spectrum remains gapless. In the ultimate limit of infinitesimally small spinon Fermi pockets, however, this assumption, although natural, has strictly speaking not been proven. We believe that an explicit calculation along the lines of Ref.~\cite{lee08} may be possible. This is left for future work.
Our focus on infinitesimally small spinon Fermi pockets allowed a controlled RG analysis, as all pertinent couplings turned out to be marginal at tree level within our RG scheme.
Our results are applicable also to systems with finite spinon Fermi pockets as long as the RG scale at which the flow diverges is large compared to the sizes of the pockets.
When the pockets become too large, further interactions can appear, that are not included in our analysis. In particular, scattering processes within the spinon Fermi pockets become possible.
If such intrapocket interactions are sufficiently attractive beyond a certain finite threshold, they could drive a Cooper pairing instability in which the U(1) gauge symmetry is broken down to a $\mathbbm Z_2$ subgroup via the Higgs mechanism~\cite{metlitski15}. On the honeycomb lattice, this stable gapped Higgs phase may be understood as being adabatically connected to the non-Abelian spin liquid ground state of the Kitaev model in an infinitesimal field~\cite{kitaev06, jiang18, hickey19}.
As the gauge-field excitations generate repulsive interactions, the \emph{weak}-coupling instability, however, will always be towards the conventional long-range-ordered state.
This suggests an unconventional deconfined quantum critical point at finite attractive intrapocket interactions between the long-range-ordered and non-Abelian Kitaev spin liquid phases.
Similar such exotic continuous quantum phase transitions between deconfined spin liquids and confined states featuring conventional magnetic~\cite{gazit18, borla20} or nonmagnetic~\cite{xu19, janssen20b} long-range orders have recently been found in square-lattice systems.
Studying the competition between Cooper and nesting instabilities at finite spinon-Fermi-pocket sizes represents an excellent direction for future research.

Recently, a number of numerical and theoretical studies have proposed spinon Fermi pocket models as effective descriptions for the field-induced intermediate phase in the antiferromagnetic Kitaev model~\cite{jiang18, jiang19, zou18, hickey19, patel19} and extensions thereof~\cite{kaib19, sodemann21}.
If these systems are indeed described within certain energy windows by spinon-Fermi-pocket models, as the numerics suggests, then we expect a low-temperature instability in the thermodynamic limit towards a long-range-ordered state.
We note, however, that finite system sizes effectively cut off the RG flow. Hence, the instability will be visible in the numerics only if the system size is larger than the scale set by the RG time at which the running interpocket coupling diverges.
The actual nature of the low-temperature order depends on microscopic properties of the system.
For the three-pocket spin liquid state with pockets at the $\boldsymbol \Gamma$ and $\mathbf K$, $\mathbf K'$ points~\cite{jiang18, jiang19}, we have argued that the proliferation of monopoles leads to plaquette or columnar VBS order on the honeycomb lattice with ordering wavevector $\mathbf Q = \mathbf K$, which coincides with the nesting vector of the noninteracting spinon Fermi surface.
When the nesting vector is different from $\mathbf K$, such as in the four-pocket state with spinon Fermi pockets at the $\boldsymbol\Gamma$ and the three $\mathbf M$ points~\cite{patel19}, the plaquette and columnar VBS orders compete with other nonmagnetic and magnetic states that feature $\mathbf Q = \mathbf M$ order. 
Which one of these will eventually be selected depends on the energetics of the microscopic system. Tuning magnetic interactions might allow one to drive the system from one order to another, yielding a possibly complex phase diagram with unconventional phase transitions.

When the extents of the spinon Fermi pockets are no longer small compared to the RG divergence scale, it is also possible that effects of imperfect nesting might become important. These could suppress the instability. For the proposed realization of the two-pocket model in the anisotropic Kitaev model in a $[111]$ magnetic field~\cite{jiang18}, for instance, the effective band mass tensor corresponding to the hole-like pocket around the $\mathbf M$ point of the hexagonal Brillouin zone will generically be anisotropic. This spoils the nesting property when the sizes of the Fermi pockets increase. By contrast, in the proposed realization of the three-pocket model in the isotropic Kitaev model~\cite{jiang18,jiang19}, the effective band masses of both the particle-like and hole-like bands near the $\boldsymbol \Gamma$ and $\mathbf K$, $\mathbf K'$ points, respectively, are isotropic. Perfect nesting therefore continues to hold as long as the spinon bands are well approximated by quadratic dispersions, see Fig.~\ref{fig:fermi-pockets}. This, in particular, will be the case in the vicinity of the proposed Lifshitz transition towards the conventional paramagnet at high fields.

We finally comment on implications for materials. The most important consequence of our work is that a gapless U(1) spin liquid with nested spinon Fermi pockets is ruled out as candidate ground state describing the magnetic behavior of any two-dimensional frustrated quantum magnet.
This applies, for instance, to the in-field behavior of $\alpha$-RuCl$_3$, for which such a state was recently proposed on the basis of the quantum oscillations observed in thermal conductivity measurements~\cite{czajka21, sodemann21}.
The assumption that this behavior indeed arises from fractionalized excitations near a neutral Fermi surface~\cite{sodemann18, czajka21} can then be reconciled with our result only if either of the following two scenarios is realized:
\paragraph*{Scenario~A.} The system features an extended spinon Fermi surface that is not nested. In this case, a gapless spin liquid that is stable up to zero temperature is possible~\cite{lee08, metlitski15}.
\paragraph*{Scenario~B.} The spin liquid state with spinon Fermi pockets describes the material only in a finite energy window. At the lower bound of this spin liquid regime, the system exhibits a finite-temperature phase transition towards a conventional long-range-ordered ground state, such as a VBS state, which spontaneously breaks lattice translational symmetries.

We note that the fact that the thermal Hall effect vanishes for field directions along Ru-Ru bonds~\cite{yokoi20}, while quantum oscillations can be observed for all measured in-plane field directions~\cite{czajka21}, has recently been argued to contradict the scenario of a simply-connected spinon Fermi surface (Scenario~A)~\cite{sodemann21}.
This may be interpreted as suggesting that Scenario~B is more likely realized in the quantum paramagnetic regime of $\alpha$-RuCl$_3$.
We believe it would be therefore be worthwhile to look for signatures of a dimerized or plaquette VBS state at ultralow temperatures in in-plane fields between 7 and 11\,T in this material.
In fact, this scenario would naturally explain three recent experimental observations:
(1)~Low-temperature thermal transport measurements show a dip at around 1\,K in the heat conductivity data as function of temperature for fixed in-plane fields in a finite field range below 11\,T~\cite{hentrich20}.
This 1\,K scale is significantly smaller than the 4\,K scale below which quantum oscillations are readily observable~\cite{czajka21}, and should therefore be expected to arise from different origin.
If Scenario~B is realized in $\alpha$-RuCl$_3$, this additional low-temperature scale might originate from the onset of long-range quantum paramagnetic order, such as VBS order, in this field range.
To test this scenario, temperature scans of the longitudinal heat conductivity at fixed in-plane fields between 7 and 11\,T down to the millikelvin regime are called for.
(2) Specific heat measurements at fixed field strength and fixed temperature below 1\,K as function of in-plane field angle reveal that characteristic deviations from the hexagonal sixfold periodicity develop in the quantum paramagnetic regime~\cite{tanaka20}. These anomalies occur for the two antiparallel in-plane field directions that are perpendicular to a particular Ru-Ru bond, indicative of a spontaneous breaking of the hexagonal lattice rotational symmetry down to a residual twofold symmetry. This behavior may be consistent with the development of dimerized staggered or zigzag VBS order.
(3)~The zero-field magnetic order in $\alpha$-RuCl$_3$ is easily melted away by a very moderate external pressure of the order of 1\,GPa or less~\cite{cui17, wang18, he18}.
At a critical hydrostatic pressure, the system exhibits a magnetic transition, along with a structural transition, towards a quantum paramagnetic state that has been understood as a dimerized VBS state~\cite{bastien18}.
This shows that VBS states strongly compete with magnetic orders in $\alpha$-RuCl$_3$, and therefore might play important roles also for zero pressure when the magnetic orders are suppressed by external fields.
Experiments in both hydrostatic pressure and finite external magnetic fields could elucidate the relation between the pressure-induced VBS state at zero field and the zero-pressure quantum paramagnet at finite fields.
On the theory side, it would be desirable to devise a pertinent microscopic model of relevance for $\alpha$-RuCl$_3$ that features zigzag order at zero field and allows fractionalized states with spinon Fermi surfaces at finite fields.
The recently proposed class of extended Kitaev-Heisenberg models~\cite{kaib19}, which can be understood as deformations of the dual version~\cite{chaloupka15} of the antiferromagnetic Kitaev model and feature a ferromagnetic Kitaev interaction~\cite{koitzsch20, sears20} and a positive off-diagonal Gamma interaction~\cite{janssen17b}, might be a useful starting point in this respect.
If a pertinent model can be found, it might allow one to tune between the cases of an extended stable spinon Fermi surface (Scenario~A) and the nested-spinon-Fermi-pocket state with its concomitant low-temperature instability (Scenario~B). 
Mapping out the corresponding phase diagram and determining thermodynamic, spectroscopic, and transport properties within the different phases and across the transitions should help to eventually clarify the true nature of the field-induced quantum paramagnetic regime in $\alpha$-RuCl$_3$.

%\paragraph*{Note added.} During the preparation of this manuscript, we became aware of a parallel work on a related model featuring a single spinon Fermi pocket, yielding, for the parts that overlap with our work, qualitatively similar results in a different RG scheme~\cite{pan21}.

%%%%%%%%%%%%%%%%%%%%%%%%%%%%%%%%%%%%
\begin{acknowledgments}
%%%%%%%%%%%%%%%%%%%%%%%%%%%%%%%%%%%%
%
Illuminating discussions with F.~Assaad, J.~Geck, S.-K.~Jian, Z.~Pan, S.~Ray, I.~Sodemann, Q.~Stahl, and H.-H.~Tu are gratefully acknowledged. 
%
%We thank S.-K.~Jian and Z.~Pan furthermore for sharing results of their work before publication.
%
This work has been supported by the Deutsche Forschungsgemeinschaft (DFG) through SFB 1143 (A07, Project No.~247310070), the W\"{u}rzburg-Dresden Cluster of Excellence {\it ct.qmat} (EXC 2147, Project No.~390858490), and the Emmy Noether program (JA2306/4-1, Project No.~411750675).
\end{acknowledgments}

%%%%%%%%%%%%%%%%%%%%%%%%%%%%%%%%%%%%
\appendix
\setcounter{equation}{0}  %%% TO FIX BUG IN REVTEX %%%
\renewcommand\theequation{A\arabic{equation}}
%%%%%%%%%%%%%%%%%%%%%%%%%%%%%%%%%%%%

\begin{widetext}

%%%%%%%%%%%%%%%%%%%%%%%%%%%%%%%%%%%%
\section*{Appendix: Details of RG calculation}
%%%%%%%%%%%%%%%%%%%%%%%%%%%%%%%%%%%%
%
In the appendix, we provide details of our RG scheme and the evaluation of loop integrals. 
We demonstrate the procedure explicitly for the two-pocket model in Eq.~\eqref{eq:ansio_theory}. 
Comments on the case of the three-pocket model are given below.
In order to perform the loop integration, we add a gauge-fixing term of the form
\begin{equation}
\mathcal{L}_{\mathrm{gf}}=-\frac{1}{2\xi}\left(\frac{1}{c^2}\partial_\tau a_\tau+\boldsymbol\nabla\cdot\mathbf{a}\right)^2,
\end{equation}
with gauge-fixing parameter $\xi \in (0,\infty)$.
For explicit computations, we use Landau gauge $\xi \to 0$.
Integrating out the fast modes with momenta in the shell $|\mathbf{p}|\in(\Lambda/b,\Lambda)$ and all frequencies $\omega \in (-\infty,\infty)$ yields the effective action for the remaining slow modes as
\begin{align}
S_< & = \int_0^{\Lambda/b}\frac{\rmd\mathbf{p}}{(2\pi)^d}\int_{-\infty}^{\infty}\frac{\rmd\omega}{2\pi}
\Bigg[
\psi_0^\dagger\left(b^{\eta_0^\omega}\rmi\omega+b^{\eta_0^{p^2}}\frac{\mathbf{p}^2}{2m_0}\right)\psi\pd_0
+\psi_{1}^\dagger\left(b^{\eta_1^\omega}\rmi\omega-b^{\eta_1^{p^2}}\frac{\mathbf{p}^2}{2m_1}\right)\psi\pd_{1}
\nonumber\\&\quad\qquad
+\frac{1}{2c^2}\left(b^{\eta_{a_\tau}/2}\mathbf{p}a_\tau-b^{\eta_{\mathbf{a}}^{\omega^2}/2}\omega\mathbf{a}\right)^2
+\frac{1}{2}b^{\eta_{\mathbf{a}}^{p^2}}\left(\mathbf{p}\times\mathbf{a}\right)^2\Bigg]
\nonumber\\&\quad
+\int_0^{\Lambda/b}\frac{\rmd\mathbf{p}_1\rmd\mathbf{p}_2}{(2\pi)^{2d}}\int_{-\infty}^{\infty}\frac{\rmd\omega_1\rmd\omega_2}{(2\pi)^2}
\Bigg[
-\frac{g+\delta g}{2m_0}(\mathbf{p}_1+2\mathbf{p}_2)\cdot\mathbf{a}\psi^\dagger_0\psi\pd_0
+\frac{g+\delta g'}{2m_1}(\mathbf{p}_1+2\mathbf{p}_2)\cdot\mathbf{a}\psi^\dagger_{1}\psi\pd_{1}
\nonumber\\&\quad\qquad
-\rmi (e + \delta e) a_\tau \psi^\dagger_0\psi\pd_0 
-\rmi (e + \delta e^\prime) a_\tau \psi^\dagger_{1}\psi\pd_{1}\Bigg]
\nonumber\\&\quad
+\int_0^{\Lambda/b}\frac{\rmd\mathbf{p}_1\rmd\mathbf{p}_2\rmd\mathbf{p}_3}{(2\pi)^{3d}}\int_{-\infty}^{\infty}\frac{\rmd\omega_1\rmd\omega_2\rmd\omega_3}{(2\pi)^3}
\Bigg[
\frac{g^2+\delta g^2}{2m_0}\mathbf{a}\cdot\mathbf{a}\psi^\dagger_0\psi\pd_0
-\frac{g^2+\delta g^{2\prime}}{2m_1}\mathbf{a}\cdot\mathbf{a}\psi^\dagger_{1}\psi\pd_{1}
+(u+\delta u)\psi^\dagger_0\psi\pd_0\psi^\dagger_{1}\psi\pd_{1}
\Bigg],
\end{align}
where in the second line we define $\mathbf{p}\times\mathbf{a} \equiv p_x a_y - p_y a_x$ in $d=2$ spatial dimensions, and $\mathbf p_1$ and $\mathbf p_2$ in the third line correspond to the momenta of the gauge field $\mathbf a \equiv \mathbf a(\omega_1, \mathbf p_1)$ and the fermion fields $\psi_i \equiv \psi_i(\omega_2, \mathbf p_2)$, $i=0,1$, respectively.
In the effective action, the fermion anomalous dimensions $\eta_0^\omega$, $\eta_0^{p^2}$, $\eta_1^\omega$, $\eta_1^{p^2}$, and gauge-field anomalous dimensions $\eta_{a_\tau}$, $\eta_{\mathbf a}^{\omega^2}$, $\eta_{\mathbf a}^{p^2}$ arise, at the one-loop order, from the fermion self-energy and polarization diagrams in Figs.~\ref{fig:diagrams}(a)--(c) and (d),(e), respectively.
The explicit vertex corrections $\delta e$, $\delta e^\prime$, $\delta g$, $\delta g^\prime$, $\delta g^2$, and $\delta g^{2\prime}$ are related to the fermion anomalous dimensions by means of the Ward identities
\begin{align}
\frac{\delta e}{e} & = \eta_0^\omega, &
\frac{\delta e'}{e} & = \eta_1^\omega, &
\frac{\delta g}{g} & = \frac{\delta g^2}{g^2} 
=  \eta_0^{p^2}, &
\frac{\delta g'}{g} & = \frac{\delta g^{2\prime}}{g^2} 
= \eta_1^{p^2},
\end{align}
which are ensured by gauge invariance.
We have explicitly verified by evaluating the corresponding diagrams in Figs.~\ref{fig:diagrams}(a)--(c) and (f)--(n) for the fermion anomalous dimensions and vertex corrections, respectively, that the above identities hold within our RG scheme to the one-loop order.
This represents an important cross-check of our calculations.
The effective action therefore remains invariant under local U(1) transformations, despite the fact that our momentum-shell regularization explicitly breaks gauge invariance.
Note that the model lacks Lorentz invariance, such that $\eta_0^\omega$ and $\eta_0^{p^2}$, as well as $\eta_1^\omega$ and $\eta_1^{p^2}$, are independent. The same is true, in principle, for $\eta_{a_\tau}$, $\eta_{\mathbf a}^{\omega^2}$, and $\eta_{\mathbf a}^{p^2}$, although these happen to vanish at the one-loop order, as discussed below.
The vertex correction $\delta u$ that renormalizes the interpocket interaction is obtained at the one-loop order from the diagrams shown in Fig.~\ref{fig:diagrams}(o)--(u).

Next, we rescale momenta as $\mathbf{p}\mapsto\mathbf{p}/b$ and frequencies as $\omega\mapsto\omega/b^z$, with $z$ denoting the dynamical critical exponent to be determined below.
Renormalizing the fields as
\begin{align}
\psi_0&\mapsto b^{(d+2z-\eta_0^\omega)/2}\psi_0, &
a_\tau&\mapsto b^{(4 + d - z - \eta_{a_\tau} + \eta_{\mathbf{a}}^{\omega^2} - \eta_{\mathbf{a}}^{p^2})/2}a_\tau, 
\displaybreak[1]\\
\psi_{1}&\mapsto b^{(d+2z-\eta_1^\omega)/2}\psi_{1}, &
\mathbf{a}&\mapsto b^{(2+d+z-\eta_{\mathbf{a}}^{p^2})/2}\mathbf{a},
\end{align}
and choosing $z=2+\eta_0^\omega-\eta_0^{p^2}$ allows us to keep the form of the noninteracting part of the effective action fixed, provided that the ratio of effective band masses $m_1/m_0$ and the speed of artificial light $c$ are renormalized as
\begin{align} \label{eq:RGflow_aniso_app_a}
\frac{d (\tfrac{m_1}{m_0})}{d\ln b}&=(2 - z + \eta_1^\omega - \eta_1^{p^2})(\tfrac{m_1}{m_0}), 
& %\displaybreak[1]\\
\frac{d c}{d\ln b} & = \frac12 (2z - 2  - \eta_\mathbf{a}^{\omega^2} + \eta_\mathbf{a}^{p^2}) c.
\end{align}
The flow of the gauge couplings can then be written as
\begin{align} \label{eq:RGflow_aniso_app_b}
\frac{d e}{d\ln b}&=\frac{1}{2}(4-d-z - \eta_{a_\tau} - \eta_{\mathbf a}^{p^2} + \eta_{\mathbf a}^{\omega^2}) e, &
\frac{d g}{d\ln b}&=\frac{1}{2}(4 - d - z - \eta_{\mathbf a}^{p^2}) g.
\end{align}
Note that diagrams with closed particle-hole loops vanish in the present type of models. This can be understood from the spinon band structure in the situation when the Fermi pockets have shrunk to isolated Fermi points [Fig.~\ref{fig:bandstructure}(b)], which does not allow particle-hole fluctuations of the same spinon flavor.
Technically, it arises from the fact that the frequency poles of the particle and hole propagators are located in the same complex half-plane, such that the integral over frequency vanishes.
This applies, for instance, to the polarization diagram in Fig.~\ref{fig:diagrams}(e), which is the only one-loop diagram that contributes to the gauge-field anomalous dimensions, implying that $\eta_{a_\tau} = \eta_\mathbf{a}^{\omega^2} = \eta_\mathbf{a}^{p^2} = 0$ at this order.
As a consequence, the electric and magnetic gauge couplings $e$ and $g$ have the same scaling dimension and the ratio $g/e$ is marginal at the one-loop order.
At higher loop orders, however, we expect finite contributions to $\eta_{a_\tau}$, $\eta_\mathbf{a}^{\omega^2}$, and $\eta_\mathbf{a}^{p^2}$, lifting the RG invariance of the ratio $g/e$.
Finally, the flow of the interpocket coupling reads
\begin{align}  \label{eq:RGflow_aniso_app_c}
\frac{d u}{d\ln b}=(z-d-\eta_0^{\omega}-\eta_1^\omega)u+\delta u.
\end{align}
Evaluating the pertinent one-loop diagrams shown in Fig.~\ref{fig:diagrams} leads to the forms displayed in Eqs.~\eqref{eq:RGflow_aniso_a}--\eqref{eq:RGflow_aniso_e}.
The functions $F$, $G$, and $H$ occurring in these equations are found in Landau gauge $\xi \to 0$ explicitly as
\begin{align}
F(x,y) & = \frac{y}{(1+2y)^4}\left[2y^2(1+8y)+4y^2(1-4y)x-(1+6y+18y^2+8y^3)x^2\right], \displaybreak[1]\\
G(x,y) & = \frac{y}{(1+2y)^4}\left[2y^2(1+8y+32y^2+16y^3)+4y^2(1+8y)x+(1+8y+22y^2+8y^3)x^2\right],\displaybreak[1]\\
H(x,y) & =
\frac{y}{16(1+2y)^5}\big[4y^4(5+50y+192y^2+320y^3+128y^4) + 16y^4(1+10y+32y^2)x
-4y^2(1+10y+30y^2-20y^3)x^2 
\nonumber\\&\quad
- 8y^2(1+10y+10y^2+4y^3)x^3
%
%\nonumber\\&\quad
%
+(3+30y+100y^2+184y^3+180y^4+72y^5)x^4\big].
\end{align}

The renormalization of the three-pocket model is carried out analogously. Due to the discrete symmetry that connects the two hole-like pockets parametrized by $\psi_1$ and $\psi_2$ in this model, both fermion fields are rescaled with the same anomalous dimension $\eta_1^\omega \equiv \eta_2^\omega$, and their effective band masses also receive the same loop correction, i.e., $\eta_1^{p^2} \equiv \eta_2^{p^2}$.
As a consequence, Eq.~\eqref{eq:RGflow_aniso_app_a}--\eqref{eq:RGflow_aniso_app_c} hold in the same form also for the three-pocket model.
For the flow of the hole-hole interpocket coupling, we similarly obtain
\begin{equation}
\frac{d \tilde{u}}{d\ln b} = (z-d-2\eta_1^\omega)\tilde{u}+\delta \tilde{u}.
\end{equation}
The explicit evaluation of the loop integrals leads to Eq.~\eqref{eq:RGflow_f} in the main text, with the function $\tilde G$ reading in Landau gauge
\begin{align}
\tilde G(x,y) & = \frac{y}{(1+2y)^4}\left[2y^2(3+24y+32y^2+16y^3)+12y^2x-(1+4y+14y^2+8y^3)x^2\right].
\end{align}
\end{widetext}

%%%%%%%%%%%%%%%%%%%%%%%%%%%%%%%%%%%%%%%%%%%%%%%%%%%%%%%%%%%%%%%%%%%%%%%
% BIBLIOGRAPHY: FOR USE WITH BIBTEX
%%%%%%%%%%%%%%%%%%%%%%%%%%%%%%%%%%%%%%%%%%%%%%%%%%%%%%%%%%%%%%%%%%%%%%%
\bibliographystyle{longapsrev4-2}
\bibliography{U1SL-Lifshitz}
%%%%%%%%%%%%%%%%%%%%%%%%%%%%%%%%%%%%%%%%%%%%%%%%%%%%%%%%%%%%%%%%%%%%%%%

\end{document}